\documentclass[aps,pra,twocolumn]{revtex4}
\pdfoutput=1
\usepackage[colorlinks=true]{hyperref}
\usepackage{graphicx}
\usepackage{amsmath}
\usepackage{amssymb}
\usepackage{color}
\usepackage{bm}

\definecolor{red}{rgb}{0.75,0,0}
\definecolor{blue}{rgb}{0,0,0.75}
\definecolor{green}{rgb}{0,0.5,0}

\DeclareMathOperator{\cn}{cn}
\DeclareMathOperator{\am}{am}

\begin{document}
\title{Softly Constrained Films}

\author{Luca Giomi}
\email{lgiomi@sissa.it}
\affiliation{SISSA, International School for Advanced Studies, Via Bonomea 265, 34136 Trieste, Italy}

\date{\today}

\begin{abstract}
	
The shape of materials is often subject to a number of geometric constraints that limit the size of the system or fix the structure of its boundary. In soft and biological materials, however, these constraints are not always hard, but are due to other physical mechanisms that affect the overall force balance. A capillary film spanning a flexible piece of wire or a cell anchored to a compliant substrate by mean of adhesive contacts are examples of these softly constrained systems in the macroscopic and microscopic world. In this article I review some of the important mathematical and physical developments that contributed to our understanding of shape formation in softly constrained films and their recent application to the mechanics of adherent cells.
	
\end{abstract}

\maketitle

\section{\label{sec:introduction}Introduction}

A typical problem in the mechanics of soft materials consists in finding the optimal shape of a system in the liquid phase subject to a number of geometric constraints. The problem occurs on the macroscopic scale, when one would like to determine the shape of a droplet wetting a surface \cite{DeGennes:2004} or that of a soap film spanning a rigid wire frame \cite{Isenberg:1992}. In the microscopic world, the mechanics of membranes, vesicles or micro-droplets revolves around finding the shape the minimizes the energy of the system given its volume \cite{Seifert:1997,Yang:1999}. That on the volume (or the area in two dimensions) is a \emph{hard constraint}, that results from the incompressibility of matter in the liquid phase. Similarly, imposing the shape of the boundary spanned by a two-dimensional interface, leads to vast class of constrained shape optimization problems in capillarity \cite{Oprea:2000}. It is then natural to ask how the problem would changed if the volume, area or boundary constraint was relaxed. 

Soap films bounded by elastic rods provide a table top realization of these {\em softly constrained} systems. A classic demonstration, when familiarizing with the concepts of surface tension, consists of dipping a rigid wire frame into soapy water and let the soap film form a surface of zero mean curvature (i.e. a minimal surface) \cite{Isenberg:1992}. How would the shape of the film change if the bounding frame was not rigid, but rather made of an inextensible but flexible elastic rod? And what if the boundary was made of a viscoelastic layer whose length itself was softly constrained by other physical mechanisms such as adhesion? 


The first example of this class of problems dates back to the end of the 19th century in the work of Maurice Le\'vy \cite{Levy:1884}, who investigated the stability of an infinitely long cylindrical pipe under pressure. Because the system is invariant for translation along the longitudinal direction, all the forces act on the cross-sectional plane of the pipe and the system is formally equivalent to a two-dimensional capillary film enclosed by a flexible elastic rod. This problem drew the attention of many researchers for more than a century and was finally solved a few years ago in a completely analytical manner \cite{Vassilev:2008,Djondjorov:2011}. Although the physics involved is very simple and ultimately reduces to the competition between two antagonist forces that try to bend the boundary in opposite directions, the problem provides a tremendous example of polymorphism and multi-stability. By simply increasing a single physical parameter, the surface tension, one can access an increasingly large number of stable configurations and in the limit of infinite surface tension the system admits a countably infinite number of equilibria. 

The Euler-Plateau problem, is a generalization of Le\'vy's original problem obtained if the softly constrained film is allowed to bend in three dimensions. The problem was proposed in Ref. \cite{Giomi:2012} as a link between two of the oldest problems in geometrical physics and the calculus of variations, bringing together Euler's {\em Elastica} and Plateau's problem. The union of these leads to the question of the shape of minimal surfaces bounded by elastic lines, a new class of questions at the nexus of geometry and physics. Among biological systems the paradigm of softly constrained film was applied in tissue morphogenesis to account for the looping patterns in the developing vertebrate gut tube \cite{Savin:2011} and in various two-dimensional models of adhering cells \cite{Bischofs:2008,Bischofs:2009,Banerjee:2013}.  


This short review is organized as follows. In Sec. \ref{sec:notation} we will refresh some basic concepts of differential geometry of curves, mostly to establish notation. In Sec. \ref{sec:plane}, we will consider the classic problem of a two-dimensional film bounded by an elastic rod, we will derive the analytical solution and discuss in detail the various regimes obtained by varying the surface tension of the film. The Euler-Plateau problem for softly constrained soap films in three dimensions will be discussed in Sec. \ref{sec:ep}. In Sec. \ref{sec:cell} we will focus on the Contractile Film Model for adherent cells \cite{Banerjee:2013} and see how the notion of softly constrained films finds a natural and important application to cell mechanics. Conclusions will be drawn in Sec. \ref{sec:conclusions}.

\section{\label{sec:notation}Mathematical preliminaries and notation}

\begin{figure}
\centering
\includegraphics[width=1\columnwidth]{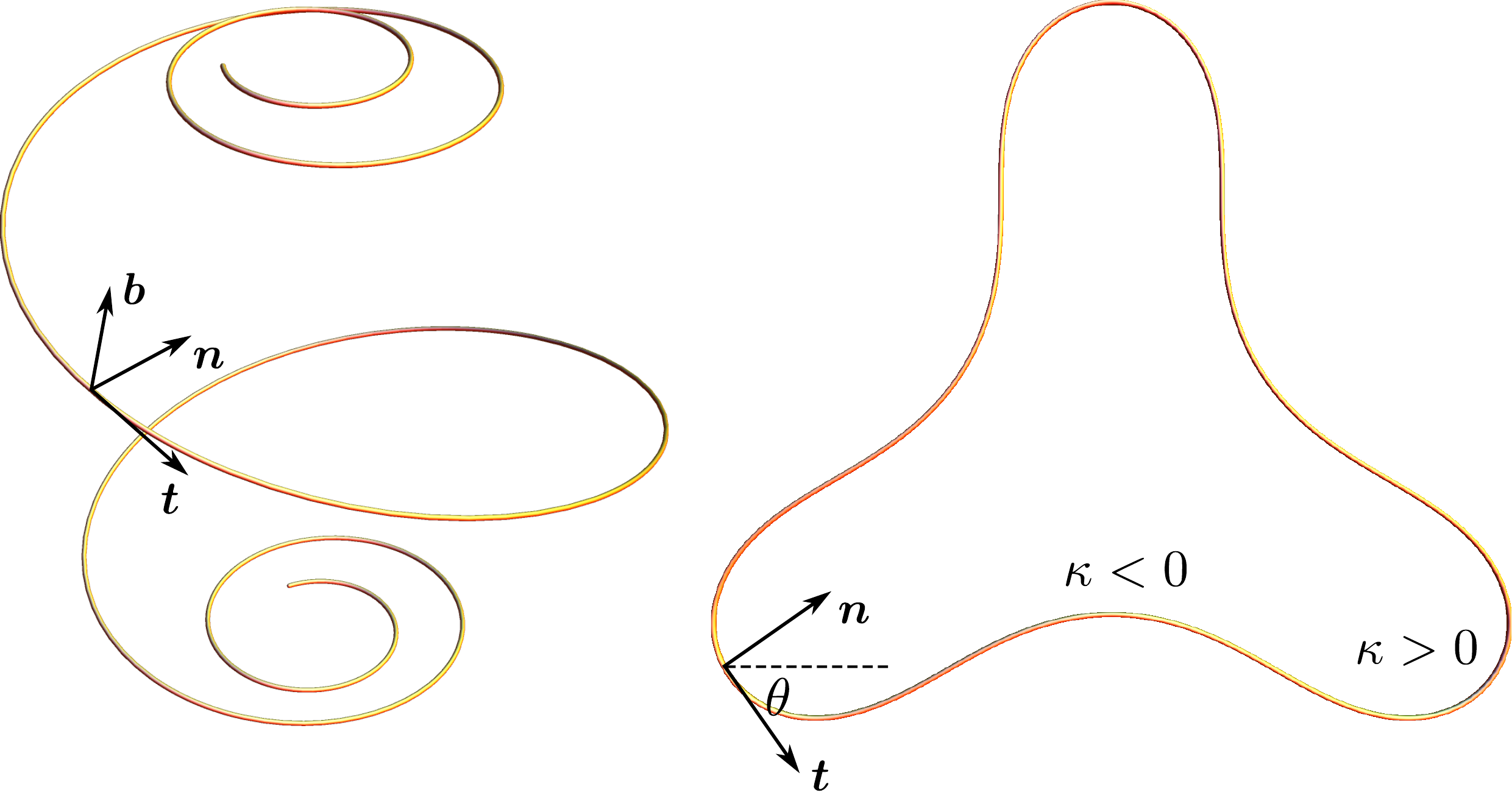}
\caption{\label{fig:frenet} The Frenet frame in three (left) and two (right) dimensions. In three dimensions the curvature $\kappa$ is conventionally taken with the positive sign and the normal vector $\bm{n}$ is directed toward the center of curvature. If a closed plane curve is oriented counterclockwise, on the other hand, $\bm{n}$ is directed toward the interior of the curve and $\kappa>0$ ($\kappa<0$) corresponds to points where the curve is convex (concave).}
\end{figure}

In all the upcoming sections, we will make a large use of the Frenet-Serret apparatus for curves in two and three dimensions, whether immersed in Euclidean space or lying on a surface. In this section we will briefly review some fundamental concepts, mostly to establish notation. Let $\bm{r}(s)$ be the position vector of a curve in $\mathbb{R}^{3}$ parametrized by its arc-length $s$. The Frenet frame of the tangent $\bm{t}=d\bm{r}/ds$, normal $\bm{n}$ and binormal vector $\bm{b}$ is described by the Frenet-Serret formulas:
\begin{subequations}\label{eq:frenet-serret}
\begin{align}
&\bm{t}'= \kappa\bm{n}\,,\\[7pt]
&\bm{n}'=-\kappa\bm{t}+\tau\bm{b}\,,\\[7pt]
&\bm{b}'=-\tau\bm{b}\,,
\end{align}
\end{subequations}
as well as the following cross-product relations:
\begin{equation}\label{eq:cross-product}
\bm{t}=\bm{n}\times\bm{b}\;,\qquad
\bm{n}=\bm{b}\times\bm{t}\;,\qquad
\bm{b}=\bm{t}\times\bm{n}\;.
\end{equation}
Here $\kappa$ and $\tau$ represents respectively the curvature and the torsion of the curve and $(\,\cdot\,)'=d(\,\cdot\,)/ds$. Eqs. \eqref{eq:frenet-serret} describe how the orthonormal frame $\{\bm{t},\bm{n},\bm{b}\}$ rotates in space as we move along the curve (Fig. \ref{fig:frenet}, left). Once $\kappa$ and $\tau$ are assigned, in the form of two differentiable functions, the fundamental theorem of space curves guarantees that the corresponding curve is uniquely determined, up to a rigid motion \cite{DoCarmo:1976}. Note that in three dimensions, the equation $\bm{t}'=\kappa\bm{n}$ does not define the sign of the curvature. That is conventionally chosen to be always positive, by incorporating the sign of $\bm{t'}$ in the definition of $\bm{n}$: i.e. $\bm{n}=\bm{t}'/|\bm{t'}|$. With this choice, the normal vector $\bm{n}$ is oriented toward the center of curvature of the curve (i.e. the center of the {\em osculating circle}). It is useful to recall that the Frenet frame is well defined only where the curvature $\kappa$ does not vanish. Where $\kappa=0$, the normal vector $\bm{n}=\bm{t}'/\kappa$ is undefined and so are $\bm{b}$ and $\tau$.

If the curve lies on a plane (Fig. \ref{fig:frenet}, right), its tangent vector can be conveniently parameterized through a single scalar function $\theta(s)$ representing the turning angle of the tangent vector $\bm{t}$. In a standard Cartesian frame: $\bm{t}=(\cos\theta,\sin\theta)$, then, using Eq. \eqref{eq:frenet-serret}, one finds $\kappa=\theta'$ and $\tau=0$. In this case, the curvature $\kappa$ identifies the curve uniquely, up to rigid motion:
\begin{equation}
\bm{r}(s) = \bm{r}(0)+\int_{0}^{s}ds'\,[\cos\theta(s')\,\bm{\hat{x}}+\sin\theta(s')\,\bm{\hat{y}}]
\end{equation}
with $\theta(s)=\theta(0)+\int_{0}^{s}ds'\,\kappa(s)$. Expressing the curvature of a plane curve as the derivative of the turning angle $\theta$, allows one to unambiguously identify the sign of $\kappa$: $\kappa>0$ ($\kappa<0$) implies that the turning angle increases (decreases) as we move along the curve, while reversing the orientation of the curve changes the sign of $\kappa$. If a closed plane curve is oriented counterclockwise and the turning angle is measured, as usual, from the $x-$axis of a Cartesian frame, $\kappa>0$ ($\kappa<0$) will then corresponds to points where the 
curve is convex (concave). This convection will be used throughout the paper.

\begin{figure}
\centering
\includegraphics[width=1\columnwidth]{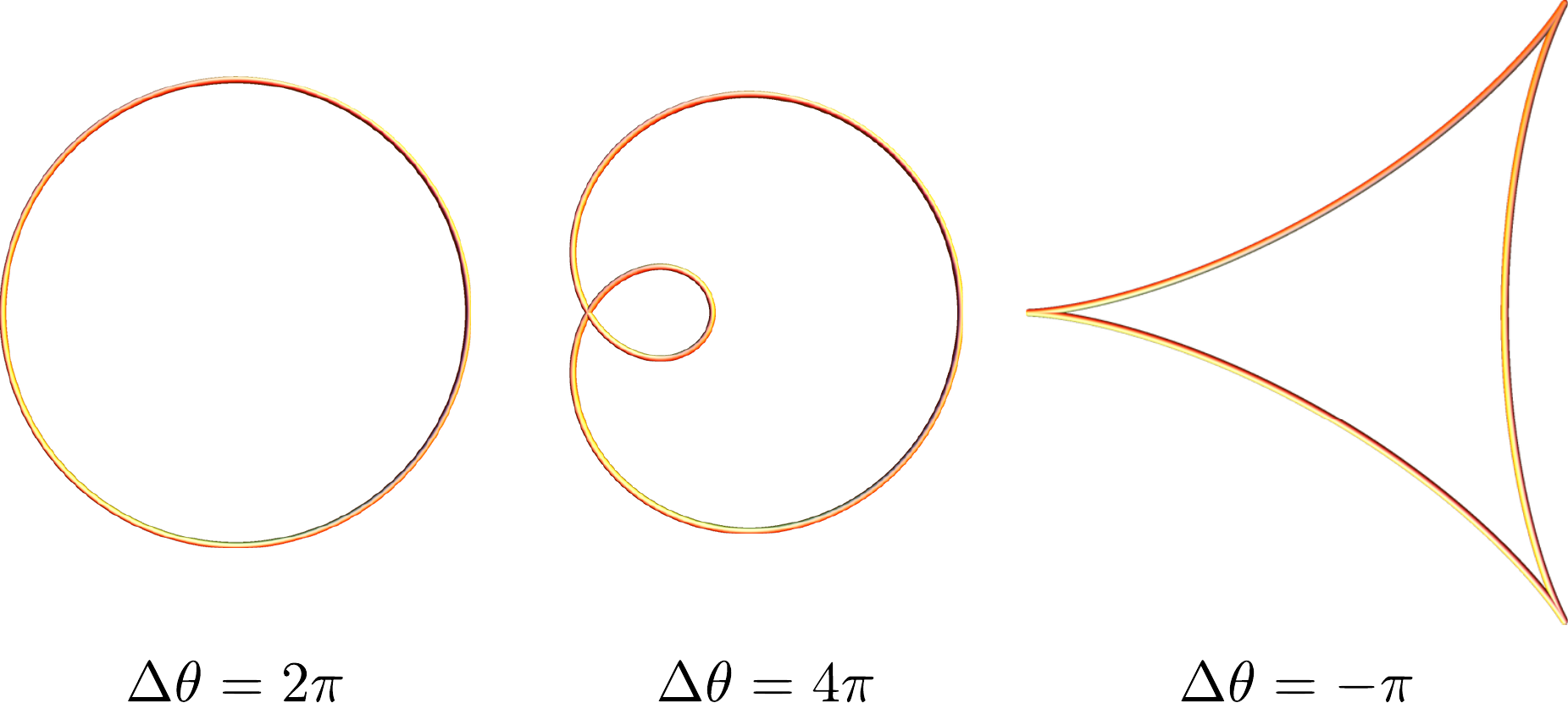}
\caption{\label{fig:turning-tangents} Three closed plane curves and their total turning angle $\Delta\theta=\oint ds\,\kappa$. (Left) In a simple closed curve, like a circle, the tangent vector rotates by $2\pi$ in one loop around the curve. (Center) In the case of a lima\c{c}on, the tangent makes two full rotations, thus $\Delta\theta=4\pi$. (Right) A deltoid curve has three cusps where the tangent vector jumps by an angle $\phi=\pi$. Eq. \eqref{eq:turning-kinks} then requires $\oint ds\,\kappa+3\pi=2\pi$, thus $\Delta\theta=-\pi$.}
\end{figure}

Closed curves, have a number of interesting global properties that will serve as important calculation tools in the later sections. The {\em four vertex theorem} \cite{DeTurck:2007}, is one of the earliest results in global differential geometry; it states that the curvature of a simple, smooth, closed curve on the plane has a least four vertices: i.e. four extrema where $\kappa'=0$ (specifically two maxima and two minima). Another fundamental property of closed plane curves is expressed by the {\em theorem of turning tangent} \cite{DoCarmo:1976,Gray:1997}. This states that:
\begin{equation}\label{eq:turning-tangents}
\oint ds\,\kappa = 2\pi m\,.
\end{equation}
The integer $m$ is called the rotation index of the curve and measures how many times the curve turns with respect to a fixed direction \cite{Gray:1997}. Simple closed curves have thus $m=1$ (Fig. \ref{fig:turning-tangents}, left), while a curve that loops twice around its center (thus self-intersects once before closing) has $m=2$ (Fig. \ref{fig:turning-tangents}, center). If a simple closed curve has kinks (singular points where the tangent vector switches discontinuously between two orientations), these will result in the total curvature as it follows:
\begin{equation}\label{eq:turning-kinks}
\oint ds\,\kappa + \sum_{i}\phi_{i} = 2\pi 	
\end{equation}
where $\phi_{i}$ is the external angle at each kink and the summation runs over all the kinks (Fig. \ref{fig:turning-tangents}, right). In the case of a convex polygon, for instance, $\kappa=0$ and \eqref{eq:turning-kinks} asserts that the sum of the external angles is equal to $2\pi$. In the case of a closed space curve, a theorem by Fenchel \cite{Fenchel:1951} states that:
\begin{equation}
\oint ds\,\kappa \ge 2\pi	
\end{equation}
where the equality holds only if the curve lies on the plane. 

A space curve that lies on a curved surface carries an important information about the local geometry its non-Euclidean habitat. This can be highlighted by introducing the Darboux frame consisting of the curve tangent vector $\bm{t}$, the surface normal vector $\bm{N}$ and the so called {\em tangent-norma}l vector (i.e. tangent to the surface and normal to the curve) $\bm{T}=\bm{N}\times\bm{t}$ \cite{Guggenheimer:1977}. The Darboux and Frenet frame can be transformed into each other through a rotation by an angle $\phi=\angle(\bm{b},\bm{N})$ about $\bm{t}$ that brings $\bm{b}$ onto $\bm{N}$ and $\bm{n}$ onto $\bm{T}$. In matrix form:
\begin{equation}\label{eq:darboux-frame}
\left(
\begin{array}{c}
\bm{t}\\
\bm{T}\\
\bm{N}	
\end{array}
\right)
= \left(
\begin{array}{ccc}
1 & 0 & 0 \\
0 & \cos\phi &  \sin\phi \\	
0 &-\sin\phi &  \cos\phi
\end{array} 
\right)
\left(
\begin{array}{c}
\bm{t} \\
\bm{n} \\
\bm{b}	
\end{array}
\right)\;,
\end{equation} 
Then, taking a derivative with respect to the arclength $s$ and using the Frenet-Serret formulas \eqref{eq:frenet-serret} one obtains the equations:
\begin{subequations}\label{eq:darboux}
\begin{align}
&\bm{t}'=\kappa_{g}\bm{T}+\kappa_{n}{\bm N}\,,\\[7pt]
&\bm{T}'=-\kappa_{g}\bm{t}+\tau_{g}\bm{N}\,,\\[7pt]
&\bm{N}'=-\kappa_{n}\bm{t}-\tau_{g}\bm{T}\,.
\end{align}
\end{subequations}
that describe how the orthonormal frame $\{\bm{t},\bm{T},\bm{N}\}$ rotates as we move along the curve. $\kappa_{g}=\kappa\cos\phi$ is the {\em geodesic curvature},  $\kappa_{n}=-\kappa\sin\phi$ is the {\em normal curvature}, while $\tau_{g}=\tau+\phi'$ is the geodeisc torsion. The geometrical meaning of $\kappa_{g}$ and $\kappa_{n}$ becomes evident by writing Eq. (\ref{eq:darboux}a) in the form:
\begin{equation}
\kappa\bm{n} = \kappa_{g}\bm{T}+\kappa_{n}\bm{N}\;.
\end{equation}
$\kappa_{n}$ and $\kappa_{g}$ are the projections of the curvature vector $\kappa\bm{n}$ along the surface normal and tangent direction respectively. Thus $\kappa_{n}$ represents the contribution to the curvature due to the fact that the underlying surface is itself curved, while $\kappa_{g}$ is the intrinsic curvature of the curve. Geodesics, being the most direct path between two points on a surface, have $\kappa_{g}=0$, while curves with $\kappa_{n}=0$ are called asymptotic lines. An alternative parametrization of the Darboux frame (which we will adopt in Sec. \ref{sec:ep}), makes use of the angle $\vartheta=\phi+\pi/2$ between the curve normal $\bm{n}$ and the surface normal $\bm{N}$. In this case: $\kappa_{g}=\kappa\sin\vartheta$ and $\kappa_{n}=\kappa\cos\vartheta$.

\section{\label{sec:plane}Planar films bounded by elastic rods}

\begin{figure}[b]
\centering
\includegraphics[width=\columnwidth]{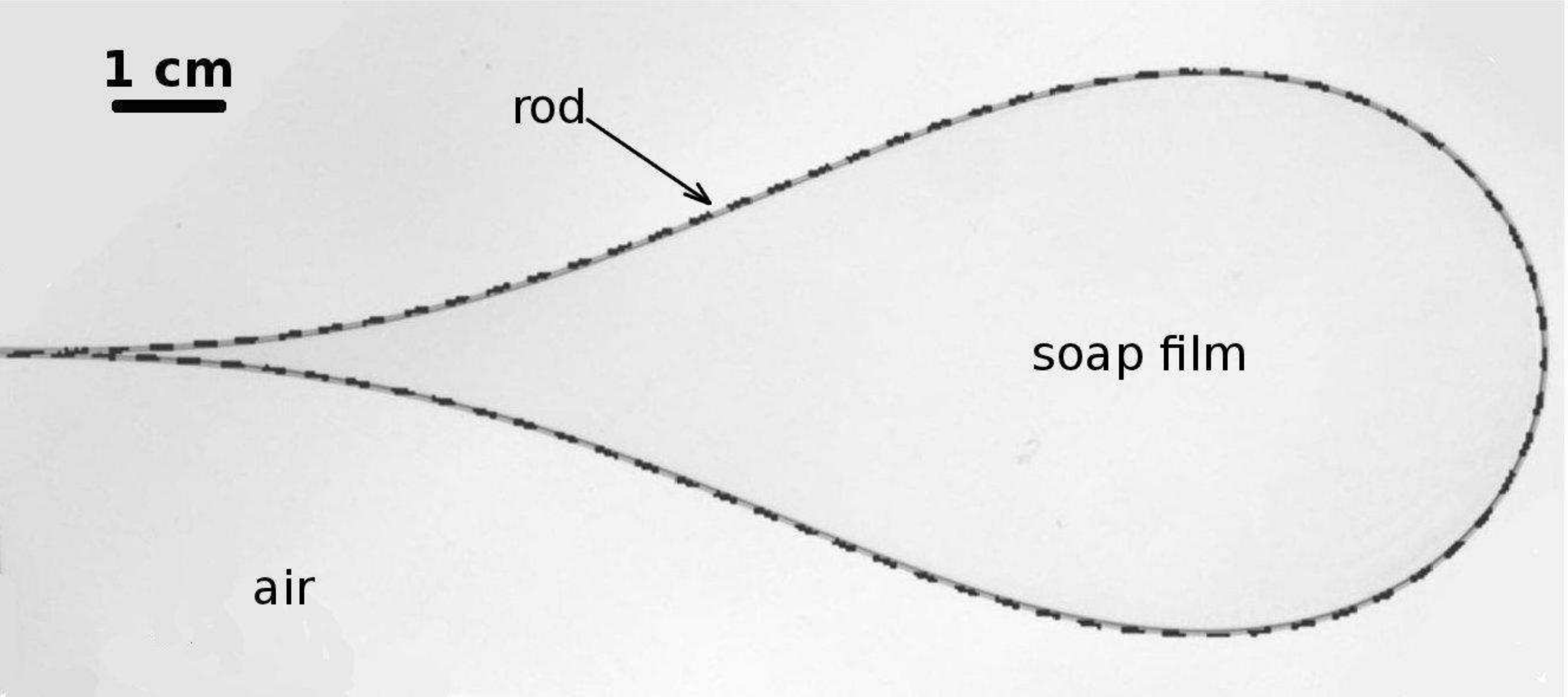}
\caption{\label{fig:mora}A rod enclosing a film of a commercial dishwashing liquid. The Young modulus of the rod is $E = 3.55 \pm 0.1$ GPa, the radius is $\rho = 0.25$ mm, and the surface tension of the film is $\sigma = 27.3$ mN/m. The dashed curve is obtained from the theoretical solution of Eq. \eqref{eq:solution} with no fitting parameter and is described in detail in Sec. \ref{sec:buckled-shapes}. Reproduced from Ref. \cite{Mora:2012}.}
\end{figure}

In Sec. \ref{sec:introduction} we introduced the problem of a soap-film bounded by an elastic rod (Fig. \ref{fig:mora}) as the prototype problem for softly constrained films. Here we review the simpler (and yet incredibly rich) version of the problem, in which the system is constraint to lie on a plane and its geometry is entirely determined by that of the boundary curve. The problem has a longstanding history as it maps one-to-one to the important problem of finding the cross sectional shape of an infinitely long cylindrical pipe subject to a pressure difference. Since the pipe is flat along its longitudinal direction, its bending energy is completely determined by the curvature of the boundary and the system is physically equivalent to a two-dimensional rod (representing the boundary of the pipe) subject to a constant normal force (i.e. the pressure difference). 

Le\'vy \cite{Levy:1884} was the first one to formulate the problem at the end of the 19th century. He investigated the stability of the circular shape and reduced the calculation of the non-circular solutions after buckling to an algebraic problem involving elliptic integrals. Later, Halphen \cite{Halphen:1888} and Greenhill \cite{Greenhill:1889} derived an exact solution in terms of Weierstrass elliptic functions and unveiled the existence of a sequence of buckling transitions to shapes of increasing order in rotational symmetry for increasing surface tension. This solution was however considered too complicated and unintuitive by many researchers who continued looking for alternative exact and approximated solutions. 

Some fifty years after the original work by L\'evy, Carrier resumed the problem, found an expression for the curvature of the buckled ring in terms of Jacobi elliptic functions and worked out an approximate solution near the onset of buckling. This analysis as been recently reconsidered by Adams \cite{Adams:2008} who further extended Carrier's work. Tadjbakhsh and Odeh \cite{Tadjbakhsh:1967}, provided a rigorous study of the boundary-value problem describing the shape of the ring and the associated variational problem. In addition, they performed the first numerical analysis of the problem and visualized a variety of post-buckling solutions, including complicated self-intersecting curves. 

Flaherty {\em et al}. \cite{Flaherty:1972} lifted the problem to a whole new level by proposing a novel scenario for the evolution of the equilibrium shape for increasing surface tension that includes self-contact. According to this picture, upon increasing the surface tension above a threshold, the ring collapses into itself. The collapse initially occurs with the appearance of an isolated point of contact, that for larger surface tension expands into a segment, giving rise to a shape reminiscent of a two-headed tennis racket. Interesting physical and mathematical aspects of the problem have been further investigated by Arreaga {\em et al}. \cite{Arreaga:2002}, Capovilla {\em et al}. \cite{Capovilla:2002} and Guven \cite{Guven:2006}. 

A completely analytical solution was finally given in a remarkable series of papers by Vassilev, Djondjorov and Mladenov for shapes before \cite{Vassilev:2008} and after collapse \cite{Djondjorov:2011}. The latter case has been recently revisited by Mora {\em et al.} \cite{Mora:2012} who additionally provided and insightful comparison between analytics and experiments (Fig. \ref{fig:mora}). In the following section, we will review some fundamental aspects of the solution and highlight some qualitative features, while remanding to the original works for a fully detailed treatment. 

\subsection{\label{sec:shape-equation}Shape equation}
Let us consider the following functional representing the total mechanical energy of the system:
\begin{equation}\label{eq:energy}
E = \sigma\int_{M}dA+\oint_{\partial M}ds\,(\alpha\kappa^{2}+\beta)\;.	
\end{equation}
Here $M$ is a mapping of a two-dimensional disk representing a film of surface tension $\sigma$, constrained to lie of the plane and bounded by a flexible beam of circular cross-section with bending rigidity $2\alpha$. The boundary $\partial M$ is thus treated as an {\em Elastica} immersed in two-dimensional Euclidean space $\mathbb{R}^{2}$, and $\beta$ is a Lagrange multiplier, analogous to line tension, that enforces inextensibility of the boundary:
\begin{equation}\label{eq:constraint}
L = \oint_{\partial M}ds\;,
\end{equation}
with $L$ the perimeter. For the two-dimensional problem discussed here, we used the convention introduced in Sec. \ref{sec:introduction}, thus the tangent vector $\bm{t}$ is oriented counterclockwise, $\bm{n}$ is directed toward the interior of the loop, so that the curve is convex where $\kappa>0$ and concave where $\kappa<0$. As a mechanical energy, Eq. \eqref{eq:energy} is amenable to various interpretations. Thinking of $\sigma$ as a hydrostatic pressure, rather than a surface tension, and to $\kappa$ as the only nonzero principal curvature of a cylinder obtained by extruding $\partial M$ along the $z-$axis, Eq. \eqref{eq:energy} is equivalent to the energy of infinitely long cylindrical pipe subject to a uniform hydrostatic pressure. Analogously, replacing the constraint on the length \eqref{eq:constraint} with a constraint on the area, with $\sigma$ playing the role of a Lagrange multiplier, and interpreting $\beta$ as a tunable line tension, Eq. \eqref{eq:energy} becomes the energy of a two-dimensional vesicle consistently with the classic framework of Canham \cite{Canham:1970} and Helfrich \cite{Helfrich:1973}. 

To derive the conditions for the extremum of the functional \eqref{eq:energy}, we will impose a small virtual deformation of the boundary, $\bm{r}\rightarrow \bm{r}+\xi\bm{n}$, and use this to calculate the corresponding change in energy and the necessary conditions for the resulting shape to be a local minimum. A linear variation of the boundary energy can be calculated with standard methods in the following form \cite{Mumford:1993}:
\begin{equation}\label{eq:elastica-variation}
\delta \int_{\partial M}ds\,(\alpha\kappa^{2}+\beta) = \oint_{\partial M} ds\,[\alpha(2\kappa''+\kappa^{3})-\beta\kappa]\xi
\end{equation}
The contribution due to surface tension, can also be readily found by expressing the enclosed area as a contour integral: $A=-(1/2)\oint ds\,\bm{r}\cdot\bm{n}$. Then, taking into account that $\delta(ds)=-\xi\kappa\,ds$ and $\delta(\bm{n})=-\xi'\bm{t}$, the area variation can be readily found to be:
\begin{equation}\label{eq:area-variation}
\delta A 
= -\oint_{\partial M} ds\,\xi\;,
\end{equation} 
consistently with intuition. Combining Eqs. \eqref{eq:elastica-variation} and \eqref{eq:area-variation} and setting the integrand to zero, we finally obtained the Euler-Lagrange equation:
\begin{equation}\label{eq:shape-equation-2d}
\alpha(2\kappa''+\kappa^{3})-\beta\kappa-\sigma = 0	
\end{equation}
Physically Eq. \eqref{eq:shape-equation-2d} represents the balance between the normal component of the stress resultant $\bm{F}$ acting a cross-section of the rod and the body force $\bm{K}=-\sigma\bm{n}$ due to surface tension. Namely:
\begin{equation}\label{eq:force-balance}
\bm{F}' = \sigma\bm{n}\;.
\end{equation}
In two dimensions, the stress resultant $\bm{F}$ can be found using the principle of virtual works \cite{Guven:2006,Giomi:2012}:
\begin{equation}\label{eq:stress-resultant-2d}
\bm{F}=(\alpha\kappa^{2}-\beta)\bm{t}+2\alpha\kappa'\bm{n}\;.
\end{equation}
Differentiating Eq. \eqref{eq:stress-resultant-2d} and using Eq. \eqref{eq:force-balance} readily gives Eq. \eqref{eq:shape-equation-2d}.

Eq. \eqref{eq:shape-equation-2d} belongs in the class of equations describing conservative systems with one degree of freedom \cite{Vassilev:2008}. Integrating Eq. \eqref{eq:shape-equation-2d} with respect to $\kappa$, gives:
\begin{equation}\label{eq:kinematic-analogy} 
\mathcal{E} = \alpha (\kappa')^{2} + U\;,
\end{equation}
where $U$ is the fourth-order polynomial:
\begin{equation}\
U=\tfrac{1}{4}\alpha \kappa^{4}-\tfrac{1}{2}\beta\kappa^{2}-\sigma \kappa\;,
\end{equation} 
and $\mathcal{E}$ an integration constant. Thus, interpreting the quantity $s$ as a time and $\kappa$ as a position, Eq. \eqref{eq:shape-equation-2d} corresponds to the equation of motion of a particle of mass $2\alpha$ and kinetic energy $\alpha(\kappa')^{2}$, moving in a potential $U$ and whose total energy $\mathcal{E}$ is conserved.

Important geometric information about the shape of this two-dimensional softly constraint film, can be found directly from Eqs. \eqref{eq:force-balance} even before integrating Eq. \eqref{eq:shape-equation-2d}. This was first established in \cite{Arreaga:2002} and later expanded in \cite{Capovilla:2002,Guven:2006}. It is not hard to verify that the normal vector $\bm{n}$ can be expressed as an arc-length derivative of a vector field:
\begin{equation}
\bm{n} = \frac{d}{ds}[(\bm{r}\cdot\bm{t})\bm{n}-(\bm{r}\cdot\bm{n})\bm{t}]\;.
\end{equation}
Thus, the force balance relation \eqref{eq:force-balance} can be recast as an identity for the plane vector:
\begin{equation}
[\alpha\kappa^{2}-\beta+\sigma(\bm{r}\cdot\bm{n})]\bm{t}+[2\alpha\kappa'-\sigma(\bm{r}\cdot\bm{t})]\bm{n} = \bm{C}\;,
\end{equation}
where $\bm{C}$ is some constant vector along the loop. This constant is unimportant as one can always choose the reference frame in such a way that $\bm{r}\cdot\bm{t}=0$ and $\bm{r}\cdot\bm{n}=-(\alpha\kappa^{2}-\beta)/\sigma$ where $\kappa'=0$, which implies $\bm{C}=\bm{0}$. For $\sigma\ne 0$, this allows one to parametrize the curve in the Frenet frame it follows:
\begin{equation}\label{eq:guven-parametrization}
\bm{r} = \frac{\bm{F}\times\bm{z}}{\sigma} = \frac{2\alpha\kappa'}{\sigma}\,\bm{t}-\frac{\alpha\kappa^{2}-\beta}{\sigma}\,\bm{n}\;,
\end{equation}
where the first equality is obtained from Eq. \eqref{eq:cross-product} by setting $\bm{z}=\bm{b}$. Because of the symmetry of the equilibrium shapes, $\oint ds\,\bm{F}=\bm{0}$, which implies that the origin of the reference frame associated with Eq. \eqref{eq:guven-parametrization} coincides with the center of mass of the curve \cite{Capovilla:2002}. Using this and \eqref{eq:kinematic-analogy}, one can further derive a remarkable polar parametrization for the curve obtained by solving Eq. \eqref{eq:shape-equation-2d}. Namely \cite{Vassilev:2008}:
\begin{equation}\label{eq:polar-parametrization}
r^{2} = \frac{4 \alpha\mathcal{E}+\beta^{2}}{\sigma^{2}}+\frac{4\alpha\kappa}{\sigma}\;.
\end{equation}
Once the spectrum of $\mathcal{E}$ is known, Eqs. \eqref{eq:guven-parametrization} and \eqref{eq:polar-parametrization} allows one to determine the local geometry of the curve without need of further integration.

\subsection{\label{sec:analytic-solution}Analytic solution in two dimensions}

In this section we will construct the analytical solution of the shape equation \eqref{eq:shape-equation-2d} and describe some its geometrical properties. The derivation presented here follows mostly that of Ref. \cite{Veerapaneni:2009}. As a first step we note that Eq. \eqref{eq:shape-equation-2d} has a trivial solution consisting of a circular disk whose radius $R$ satisfies the cubic equation:
\begin{equation}\label{eq:circle}
\alpha R^{3}+\beta R^{2}-\alpha = 0\;,
\end{equation}
for all parameter values, with $\beta$ enforcing the condition $L=2\pi R$. For large enough surface tension, however, we might expect this configuration to become unstable and the disk to buckle into a more complex shape. To test this hypothesis we analyze the stability of the disk with respect to a small periodic displacement in the radial direction $\delta R =\epsilon\sin n\phi$, where $\phi$ is the polar angle, $n$ and integer and $\epsilon$ a small amplitude. Expanding the energy \eqref{eq:energy} to second order in $\epsilon$ yields:
\begin{equation}\label{eq:stability}
E \approx E_{0}+\frac{\pi\epsilon^{2}}{2R^{3}}[\sigma R^{3}+\beta n^{2} R^{2}+\alpha(2n^{4}-5n^{2}+2)]\;,
\end{equation}
where $E_{0}=\sigma(\pi R^{2})+2\pi R(\alpha/R^{2}+\beta)$ is the energy associated with the circular shape. Here the Langrange multiplier $\beta$ is given by Eq. \eqref{eq:circle} with $R=L/2\pi$, so that $\beta=[\alpha-\sigma(L/2\pi)^{3}]/(L/2\pi)^{2}$. Replacing this in Eq. \eqref{eq:stability}, it is easy to very that the coefficient of the second order term becomes negative when:
\begin{equation}\label{eq:critical-sigma}
\frac{\sigma L^{3}}{\alpha} > 16 \pi^{3}(n^{2}-1)\;.
\end{equation}
Thus the first mode that goes unstable is that associated with an elliptical deformation with $n=2$, corresponding to the critical value $\sigma L^{3}/\alpha>48\pi^{3}$. 

In order to calculate the curvature after buckling, let us first make Eq. \eqref{eq:shape-equation-2d} dimensionless by taking $t=s/L$, $\hat{\kappa}=L\kappa$, $\hat{\beta}=\beta L^{2}/\alpha$ and $\hat{\sigma}=\sigma L^{3}/\alpha$. We then have:
\begin{equation}\label{eq:shape-equation-nodim}
2\hat{\kappa}''+\hat{\kappa}^{3}-\hat{\beta}\hat{\kappa}+\hat{\sigma}=0\;.
\end{equation}
Without loss of generality, we can choose $t=0$ as the point where the derivative of $\hat{\kappa}$ vanishes: $\hat{\kappa}'(0)=0$. In addition, we can take:
\begin{equation}\label{eq:boundary-conditions}
\hat{\kappa}(0)=\hat{\kappa}(1)=\hat{\kappa}_{0}\;,
\end{equation}
and determine the constant $\hat{\kappa}_{0}$ later on by using the fact that the curve is closed, thus its integral must satisfy the theorem of turning tangents \eqref{eq:turning-tangents}. Integrating Eq. \eqref{eq:shape-equation-nodim} over $\hat{\kappa}$ and using Eq. \eqref{eq:boundary-conditions}, we obtain:
\begin{equation}
(\hat{\kappa}')^{2}+\tfrac{1}{4}(\hat{\kappa}^{4}-\hat{\kappa}_{0}^{4})-\tfrac{1}{2}\hat{\beta}(\hat{\kappa}^{2}-\hat{\kappa}_{0}^{2})-\hat{\sigma}(\hat{\kappa}-\hat{\kappa}_{0})=0\;.
\end{equation}
Introducing the new variable $y=1/(\hat{\kappa}_{0}-\hat{\kappa})$, we can reduce the order of the nonlinearity by one unit:
\begin{equation}
(y')^{2} 
= (\hat{\kappa}_{0}^{3}-\hat{\beta}\hat{\kappa}_{0}-\hat{\sigma})y^{3}
+ (\tfrac{1}{2}\hat{\beta}-\tfrac{3}{2}\hat{\kappa}_{0}^{2})y^{2}
+\hat{\kappa}_{0}y-\tfrac{1}{4}\;.
\end{equation}
This equation is of the form:
\[
(y')^{2}=P(y)=h^{2}(y-a)(y-b)(y-c)\;,
\]
with $h^{2}=\hat{\kappa}_{0}^{3}-\hat{\beta}\hat{\kappa}_{0}-\hat{\sigma}$ and $a$, $b$ and $c$ the roots of the cubic polynomial $P(y)$, and is suitable to be solved in terms of elliptic functions. Next we make the assumption that $P(y)$ has a single real root $y=\alpha$ and a pair of complex conjugate roots $y=\beta\pm i\gamma$. Thus $P(y)$ admits the factorization:
\begin{equation}
P(y) = h^{2}(y-\alpha)[(y-\beta)^{2}+\gamma^{2}]\;,	
\end{equation}
which allows us to calculate the elliptic integral \cite{Greenhill:1892}:
\begin{equation}\label{eq:inverse-solution}
t = \int_{y}^{\infty}\frac{dy}{\sqrt{P(y)}} = \omega^{-1}\cn^{-1}\left(\frac{y-z_{1}}{y-z_{2}},m\right)\;,
\end{equation}
where $z_{1}$ and $z_{2}$ are the roots of the quadratic equation:
\begin{equation}\label{eq:quadratic}
z^{2}-2\alpha z+2\alpha\beta-(\beta^{2}+\gamma^{2}) = 0\;,
\end{equation}
and $\omega$ and $m$ are given by:
\begin{equation}
\omega^{2}=\frac{h(z_{1}-z_{2})}{2}\;,\qquad
m^{2}=\frac{\beta-z_{2}}{z_{1}-z_{2}}\;.
\end{equation}
Finally, solving Eq. \eqref{eq:inverse-solution} for $y$ and going back to our original variables, we have:
\begin{equation}\label{eq:solution}
\hat{\kappa}=\hat{\kappa}_{0}-\frac{1-\cn(\omega t,m)}{z_{1}-z_{2}\cn(\omega t,m)}\;.
\end{equation}
In Eqs. \eqref{eq:inverse-solution} and \eqref{eq:solution} we use the standard notation for Jacobi elliptic function \cite{Davis:2010}. Namely, given the incomplete elliptic integral of the first kind:
\[
u = F(\phi,m) = \int_{0}^{\phi}\frac{dt}{\sqrt{1-m^{2}\sin^{2}t}}\;,
\]
with $0<m^{2}<1$ the elliptic modulus, then $\phi$ is the Jacobi amplitude: $\phi=\am(u,m)$ and $\cn(u,m)=\cos\phi$. As it was noted in \cite{Veerapaneni:2009}, starting from the assumption that polynomial $P(y)$ has three real roots instead of one, would have led to a contradiction. Indeed, the fact that $P(y)$ has three real roots implies that there are four {\em distinct} values of $\hat{\kappa}$ for which $\hat{\kappa}'=0$. The solution obtained by assuming three real roots has however only two distinct extrema: hence a contradiction. 

The solution \eqref{eq:solution} satisfies by construction the boundary condition $\hat{\kappa}(0)=\hat{\kappa}_{0}$ and $\hat{\kappa}'(0)=0$. In order for it to be a legitimate solution of the problem, we further need it to be periodic, so that $\hat{\kappa}(0)=\hat{\kappa}(1)$ and satisfy the theorem of turning tangents \eqref{eq:turning-tangents}. Periodicity can be easily implemented by recalling that $\cn(x+4nK(m))=\cn(x)$, where $K(m)=F(\frac{\pi}{2},m)$ is the complete elliptic integral of the first kind and $n$ an integer. This results in the following condition for the frequency $\omega$:
\begin{equation}\label{eq:periodicity}
\omega=4nK(m)\;.
\end{equation}
Now, the number of extrema of the function \eqref{eq:solution} in the interval $0\le t < 1$ is given by $2n$, thus by virtue of the four-vertex theorem (Sec. \ref{sec:notation}), $n \ge 2$. In the next section we will see how different values of $n$ give shapes with a different number of lobes and that all of them are local minima of the energy \eqref{eq:energy}. From the theorem of turning tangents \eqref{eq:turning-tangents} for a simple closed curve, on the other hand, we obtain:
\begin{multline}\label{eq:closure}
\hat{\kappa}_{0}-2\pi=\\
\frac{1}{z_{1}z_{2}K(m)}\left[z_{1}K(m)-\frac{z_{1}-z_{2}}{\sqrt{1-m^{2}}}\,\Pi\left(\frac{z_{2}^{2}}{z_{1}^{2}}\bigg|\frac{m^{2}}{m^{2}-1}\right)\right]\;,
\end{multline}
where $\Pi$ is the complete elliptic integral of the third kind:
\begin{equation}
\Pi(n|m) = 
\int_{0}^{\frac{\pi}{2}}\frac{dt}{(1-n\sin^{2}t)\sqrt{1-m\sin^{2}t}}\;.
\end{equation}
Solving simultaneously the transcendental equations \eqref{eq:periodicity} and \eqref{eq:closure}, the quadratic equation \eqref{eq:quadratic} and its associated cubic, allows to calculate $\hat{\kappa}_{0}$ and $\hat{\beta}$. The solution is then complete. 

\begin{figure*}
\centering
\includegraphics[width=0.9\textwidth]{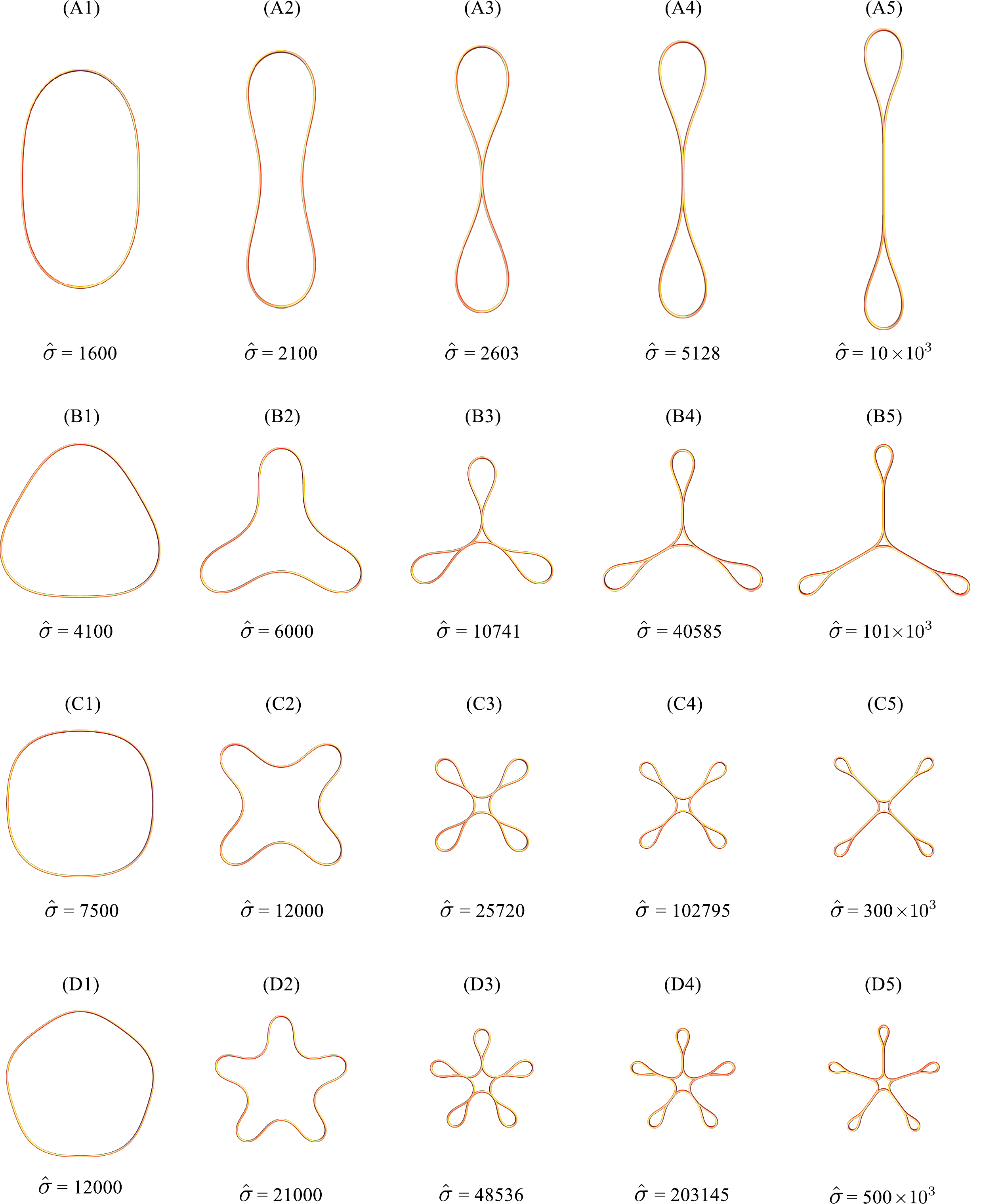}	
\caption{\label{fig:gallery}Gallery of shapes obtained by solving Eq. \eqref{eq:shape-equation-2d} for various surface tension values and with rotational symmetry index $2 \le n \le 5$. The curves in the third column show the onset of point-contact at $\sigma=\sigma_{{\rm c}n}$ (compare with Table \ref{tab:contact}). The curves in the fourth column, on the other hand, show the onset of line-contact at $\sigma=\sigma_{0n}$. In this case the curvature vanishes wherever the curve is in contact with itself.}
\end{figure*}

\subsection{\label{sec:buckled-shapes}Buckled shapes}

With the help of the machinery developed in the previous section, we can now explore the variety of shapes that can be accessed by the system upon chancing the surface tension. In Sec. \ref{sec:analytic-solution} we have shown that for any $n \ge 2$, there is a critical value $\hat{\sigma}_{{\rm b}n}=16\pi^{3}(n^{2}-1)$ such that for $\hat{\sigma}>\hat{\sigma}_{{\rm b}n}$ the circular shape is unstable to a buckled shape having $n-$fold rotational symmetry. A theorem by Tadjbakhsh and Odeh \cite{Tadjbakhsh:1967} guarantees that for any $n\ge 2$ and $\hat{\sigma}>\hat{\sigma}_{{\rm b}n}$ there is a unique buckled solution. This is given by Eq. \eqref{eq:solution}, with the parameters $z_{1}$ and $z_{2}$ (thus $\omega$ and $m$) given by Eqs. \eqref{eq:periodicity} and \eqref{eq:closure}. Some example of these buckled shapes are given in Fig. \ref{fig:gallery} for various $n$ and $\hat{\sigma}$.

After buckling, increasing the surface tension results in a reduction of the area enclosed by the rod, with the convex regions of the curve moving away from the center of mass and the convex regions shifting toward the interior, consistently with Eq. \eqref{eq:polar-parametrization}. This trend continues until one reaches a second $n-$dependent critical surface tension $\hat{\sigma}_{{\rm c}n}$ at which two or more points on the opposite sides of the boundary come into contact. $\hat{\sigma}_{{\rm c}n}$ is called the {\em contact surface-tension} (or contact pressure) and was first introduced by Flaherty and coworkers \cite{Flaherty:1972}. To gain insight about this regime one can consider the parametrization given in Eq. \eqref{eq:guven-parametrization} in the center of mass frame. Symmetry requires that a point of contact $\bm{r}\cdot\bm{n}=0$, thus from Eq. \eqref{eq:guven-parametrization} we find that the curvature at a point of contact is given by:
\begin{equation}\label{eq:point-of-contact}
\hat{\kappa}(t_{\rm c}) = -\sqrt{\hat{\beta}}\;,
\end{equation}
where $t_{\rm c}=s_{\rm c}/L$ is the rescaled arc-length coordinate of a point of contact and the choice of the minus comes from the fact that contact occurs in the concave regions of the rod. In the physically relevant case of $n=2$, contact occurs at a minimum of the curvature thus: $\bm{F}(t_{c})=\bm{0}$ and $\hat{\kappa}(t_{c})=\hat{\kappa}_{0}$. Then, using Eq. \eqref{eq:point-of-contact} and solving Eqs. \eqref{eq:periodicity} and \eqref{eq:closure} for $\hat{\kappa}_{0}$ and $\hat{\sigma}$ we get $\hat{\kappa}_{0}=2\pi$ and $\hat{\sigma}_{c2}=2603$. It is worth to stress that at the point of contact, the curvature of the two-lobbed shape matches that of the circular ring before buckling. 

The same mechanism occurs for $n\ge 3$, in this case, however, contact does not occur at the point of minimal curvature, thus $\hat{\kappa}(t_{c})\ne\hat{\kappa}_{0}$. In order to find the contact surface tension $\hat{\sigma}_{\rm c}$ one needs to look for a solution of the Frenet-Serret equations associated with the Eq. \eqref{eq:solution} that contains two points points of arc-length $s_{\rm c}<s'_{\rm c}<L/n$ at which $\bm{r}(s_{\rm c})=\bm{r}(s'_{\rm c})$ and $\bm{t}(s_{\rm c})=-\bm{t}(s'_{\rm c})$. This task was accomplished in Ref. \cite{Tadjbakhsh:1967,Flaherty:1972} using a shooting method. An alternative and simpler method, was introduced by Djondjorov {\em et al}. \cite{Djondjorov:2011} and consists of solving Eqs. \eqref{eq:periodicity} and \eqref{eq:closure} together with a further equation for the turning angle at the point of contact. Some values of $\hat{\sigma}_{{\rm c}n}$ calculated with this method are tabulated in Table \ref{tab:contact}.

\begin{table}[t]
\begin{ruledtabular}
\begin{tabular}{ccccccc}
$n$ & 2 & 3 & 4 & 5 & 6 \\
\hline
$\hat{\sigma}_{{\rm c}n}$ & 2603.0 & 10740.6 & 25719.8 & 48535.5 & 79910.4 \\
$\hat{\sigma}_{0n}$ & 5127.7 & 40584.9 & 102795.4 & 203145.1 & 349245.2 \\
\end{tabular}
\end{ruledtabular}
\caption{\label{tab:contact}The surface tension $\hat{\sigma}_{{\rm c}n}$ and $\hat{\sigma}_{0n}$ at the onset of point and line contact for $n \le 6$. The units used here differ from those used in Refs. \cite{Flaherty:1972,Djondjorov:2011}. These and can be recovered by dividing the values reported in the table by $16\pi^3$.}
\end{table}

As it was explained by Flaherty {\em et al}. \cite{Flaherty:1972} and experimentally confirmed in a beautiful work by Mora {\em et al}. \cite{Mora:2012}, as $\hat{\sigma}$ is increases above $\hat{\sigma}_{{\rm c}n}$, the opposite sides remain in contact at one pair of points in each lobe. Simultaneously, the stress resultant $\bm{F}(t_{c})$ increases and the curvature decreases until it becomes zero at certain surface tension $\hat{\sigma}_{0n}>\hat{\sigma}_{{\rm c}n}$. From Eq. \eqref{eq:point-of-contact} this also implies that $\hat{\beta}=0$ at $\hat{\sigma}=\hat{\sigma}_{0n}$. Fig. \ref{fig:mora} shows the experimental realization by Mora {\em et al}. \cite{Mora:2012} of the self-contacting lobe together with the theoretical prediction (dashed line).  

\begin{figure}[t]
\centering
\includegraphics[width=0.8\columnwidth]{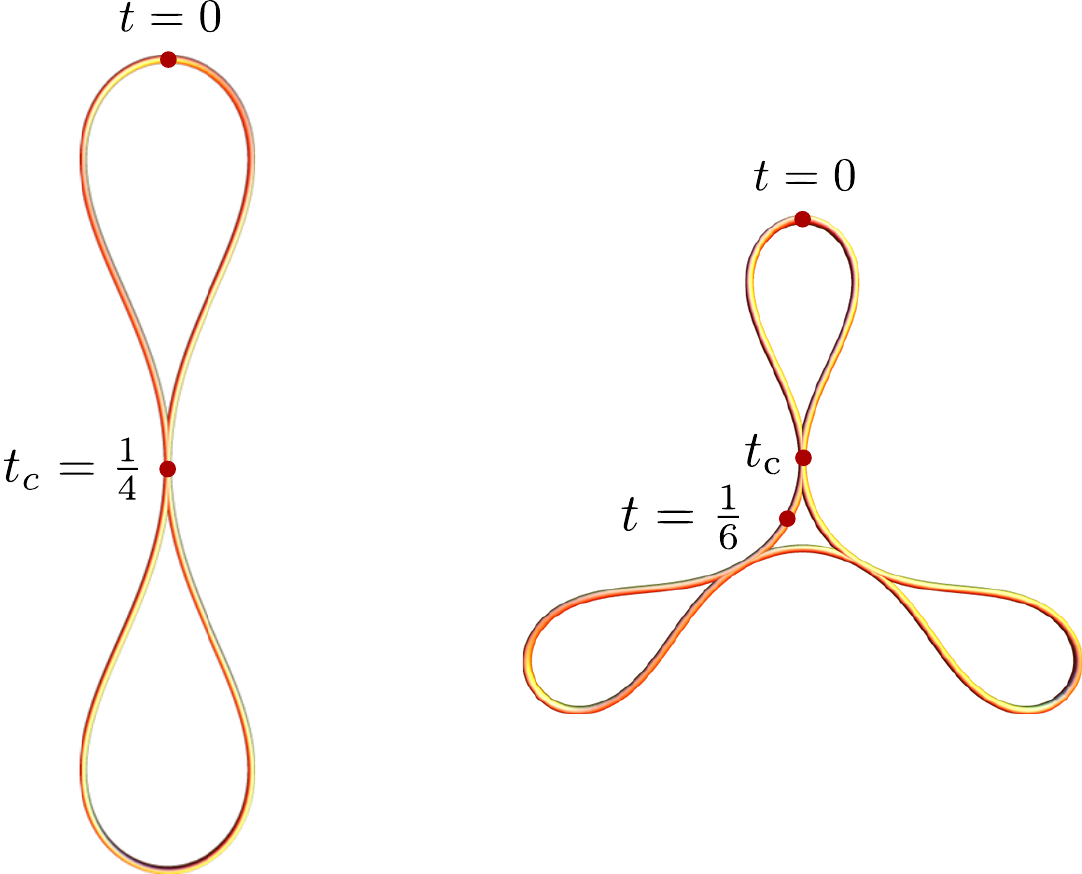}	
\caption{\label{fig:contact}Geometry at the onset of contact. In the case of the $2-$fold deformation mode, contact occurs in the center of mass of the curve, thus $t_{\rm c}=s_{\rm c}/L=1/4$ if we set $t=0$ at the tip of a lobe, where the curvature is maximal. For the generic $n-$fold mode (with $n\ge 3$), contact occurs before the curvature minimum at $t=1/n$. Thus calculating the curvature requires to separate a half-lobe in two intervals, $0\le t \le t_{\rm c}$ and $t_{\rm c}\le t \le 1/n$, and solve two separate problems with the matching condition given in Eqs. \eqref{eq:sigma0n}.}
\end{figure}

For the simplest case of $n=2$, both the critical surface tension $\hat{\sigma}_{02}$ and the shape of the lobes can be found straightforwardly by setting $t=0$ at the tip of a lobe (where the curvature is maximal) and $t_{\rm c}=1/4$ (Fig. \ref{fig:contact}). Next we can look for a function of the form given in Eq. \eqref{eq:solution} that satisfies:
\begin{equation}\label{eq:sigma02}
\hat{\kappa}\left(\tfrac{1}{4}\right) = 0\;,\qquad
\Delta\theta=\int_{0}^{\frac{1}{4}} dt\,\hat{\kappa} = \frac{\pi}{2}\,.
\end{equation}
The second condition reflects the fact that the tangent vector rotates by $\pi$ in one counterclockwise loop of the lobe, thus it rotates by $\pi/2$ if we start from the mid-point at $t=0$. Solving numerically Eqs. \eqref{eq:sigma02} for $\hat{\kappa}_{0}$ and one finds $\hat{\kappa}_{0}=28.8164$ and the value of $\hat{\sigma}_{02}$ reported in Table \ref{tab:contact}. 

For $n\ge 3$ contact does not occur at the center of the ring, thus finding $\hat{\sigma}_{0n}$ and the shape of the rod requires to introduce further equations to determine the position $t_{\rm c}$ of the contact point for each lobe. As it was suggested in Ref. \cite{Flaherty:1972} this problem can be addressed by dividing the length of a half-lobe in two intervals: from the tip of the lobe to the point of contact, $0\le t \le t_{\rm c}$, and form the point of contact to the end of the lobe where the curvature is minimal and the curve is closer to its center of mass, $t_{\rm c}\le t \le 1/n$ (Fig. \ref{fig:contact}). In each of these intervals we can then search for a solution that satisfies:
\begin{subequations}\label{eq:sigma0n}
\begin{gather}
\hat{\kappa}(t_{\rm c})=\hat{\kappa}'(0)=0\;,\qquad 
\int_{0}^{t_{\rm c}}dt\,\hat{\kappa} = \frac{\pi}{2}\;. \\[5pt]
\hat{\kappa}(t_{\rm c})=\hat{\kappa}'(1/n)=0\;,\qquad
\int_{t_{\rm c}}^{\frac{1}{n}}dt\,\hat{\kappa} = \frac{\pi}{n}-\frac{\pi}{2}\;.
\end{gather}	
\end{subequations}
Once again, we can use the general solution \eqref{eq:solution} to calculate the integrals, but in the second interval this should be transformed in such a way the derivative of the curvature vanishes at the end of the lobe, thus: $t\rightarrow 1/n-t.$ (this corresponds to translate the origin to the end of the lobe and invert the orientation). Thanks to the availability of an analytical expression for the curvature, solving Eqs. \eqref{eq:sigma0n} numerically is not hard and one can readily find the surface tension values reported in Table \ref{tab:contact} as well as the shapes shown in Fig. \ref{fig:gallery} (fourth column). For a reason that will be explained below, all the lobes constructed in this way are similar regardless of $n$, while it is only the shape of the central loop that changes becoming more and more convex as $n$ increases.

Upon increasing the surface tension above $\hat{\sigma}_{0n}$, the points of contact becomes lines of increasing length and the process continues until, in the limit of infinite $\hat{\sigma}$, the shape is completely collapsed into a star consisting of $n$ lines segments meeting in the center and regularly separated by an angle of $2\pi/n$. As it was observed in Refs. \cite{Flaherty:1972,Djondjorov:2011}, the shape of the partially collapsed loops can be constructed by rescaling that at $\hat{\sigma}_{0n}$. Eq. \eqref{eq:shape-equation-2d} is indeed invariant under the following scaling transformation:
\begin{equation}\label{eq:similarity}
(s,\kappa,\beta,\sigma) \rightarrow \left(\lambda s,\,\frac{\kappa}{\lambda},\,\frac{\beta}{\lambda^{2}},\,\frac{\sigma}{\lambda^{3}}\right)\;.
\end{equation}
Consequently, the shape of the end-loop obtained for $\hat{\sigma}>\hat{\sigma}_{0n}$ is similar to that at $\hat{\sigma}_{0n}$ up to a scaling factor $\lambda=(\hat{\sigma}_{0n}/\hat{\sigma})^{1/3}<1$. Accordingly, the length and the area  of the rescaled shape are rispectivelly $\lambda L$ and $\lambda^{2}A$, with $A$ the area of the reference shape obtained at $\hat{\sigma}_{0n}$. This beautiful geometric property can be translated into the following algorithm to construct shapes with extended regions of contact \cite{Flaherty:1972,Djondjorov:2011,Mora:2012}: {\em 1)} given $\hat{\sigma}_{0n}$ and a surface tension $\hat{\sigma}>\hat{\sigma}_{0n}$ we construct a reference shape associated with $\hat{\sigma}_{0n}$ and calculate the scaling factor $\lambda$; {\em 2)} we then rescale the reference curve so that its length is $\lambda L$; {\em 3)} finally we replace each isolated point of contact by a line segment of length: 
\begin{equation}\label{eq:contact-length}
\ell = \frac{L}{2n}\,(1-\lambda)\;,
\end{equation}
so that the total length is once again equal to $L$. Some example of shapes constructed with this procedure are shown in Fig. \ref{fig:gallery} (fifth column). It should be stressed that in the construction given here, the length of the contact lines is shorter by a factor $1/2$ with respect to that given in Ref. \cite{Djondjorov:2011}. The reason of this discrepancy is the following: if we think of the boundary curve as a physical rod, then the segments of line connecting the lobes with the central loop consists of {\em two} portions of rod in contact with each other and their length should be double counted. Hence the factor $1/2$ included in Eq. \eqref{eq:contact-length}.

\begin{figure}[t]
\centering
\includegraphics[width=0.8\columnwidth]{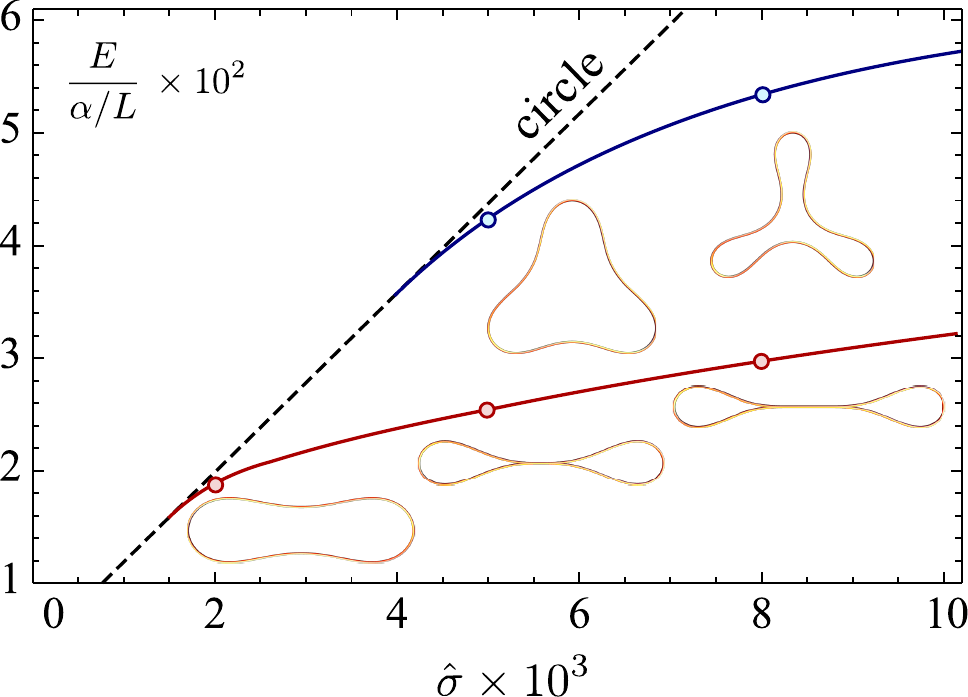}	
\caption{\label{fig:energy}The rescaled energy \eqref{eq:energy} as a function of $\hat{\sigma}$ for the $2-$ and $3-$fold symmetric solution of the shape equation \eqref{eq:shape-equation-2d}. The curves in the inset correspond to the values of $\hat{\sigma}$ marked by the dots. The dashed-line shows the energy of the circular solution $E_{0}/(\alpha/L)=\hat{\sigma}/(4\pi)+4\pi^{2}$.}
\end{figure}

Fig. \ref{fig:energy} shows a plot of the energy \eqref{eq:energy} rescaled by a factor $\alpha/L$ for two families of solutions of the shape equation \eqref{eq:shape-equation-2d}, together with some representative configurations. As one can see, solutions of larger $n$ are always energetically more expensive than the simple two-lobbed shape, thus are never found in a typical experimental realization of the problem \cite{Mora:2012}. It appears feasible, however, to take advantage of the initial configuration of the ring to bias the system toward a shape having $n\ge 3$ and possibly use the various buckling instabilities discussed here to achieve conformational changes, in the same spirit of Ref. \cite{Shim:2012}.

\subsection{Final remarks}

If the rod is allowed to intersect, increasing the surface tension above $\hat{\sigma}_{{\rm c}n}$ produces various self-intersecting curves one can also describe using Eqs. \eqref{eq:solution}, \eqref{eq:periodicity} and \eqref{eq:closure}. This solutions, however, do not have a physical relevance and for this reason they will not be discussed here. Interested readers will find a thorough discussion of this regime in Ref. \cite{Arreaga:2002}. 

Regardless of the applied interest to situations involving vessels subject to a uniform pressure (thus including the circulatory system), the problem reviewed in this section has some remarkable features. Although the physics involved is very simple and ultimately reduces to the competition between two antagonist forces that try to bend the rod in opposite directions, the simple composite structure described by the energy \eqref{eq:energy} provides a tremendous example of polymorphism and multi-stability. By simply increasing a single physical parameter, the surface tension, one can access an increasingly large number of stable configurations and in the limit $\sigma\rightarrow\infty$ the system admits a countably infinite number of minima corresponding to fully collapsed stars having an arbitrary integer number of rays departing from the center. In addition, the problem provides a fascinating example of contact and cusp geometry. Finally, Eq. \eqref{eq:shape-equation-2d} can be solved in a completely analytical manner (proviso solving numerically some transcendental equation involving elliptic integrals) and the solution is in excellent agreement with the data (see Fig. \ref{fig:mora}). In Sec. \ref{sec:cell} we will see how the techniques and the intuition built up here will turn into an useful set of theoretical tools to address an important problem in cell mechanics.

\begin{figure}[t]
\centering
\includegraphics[width=1\columnwidth]{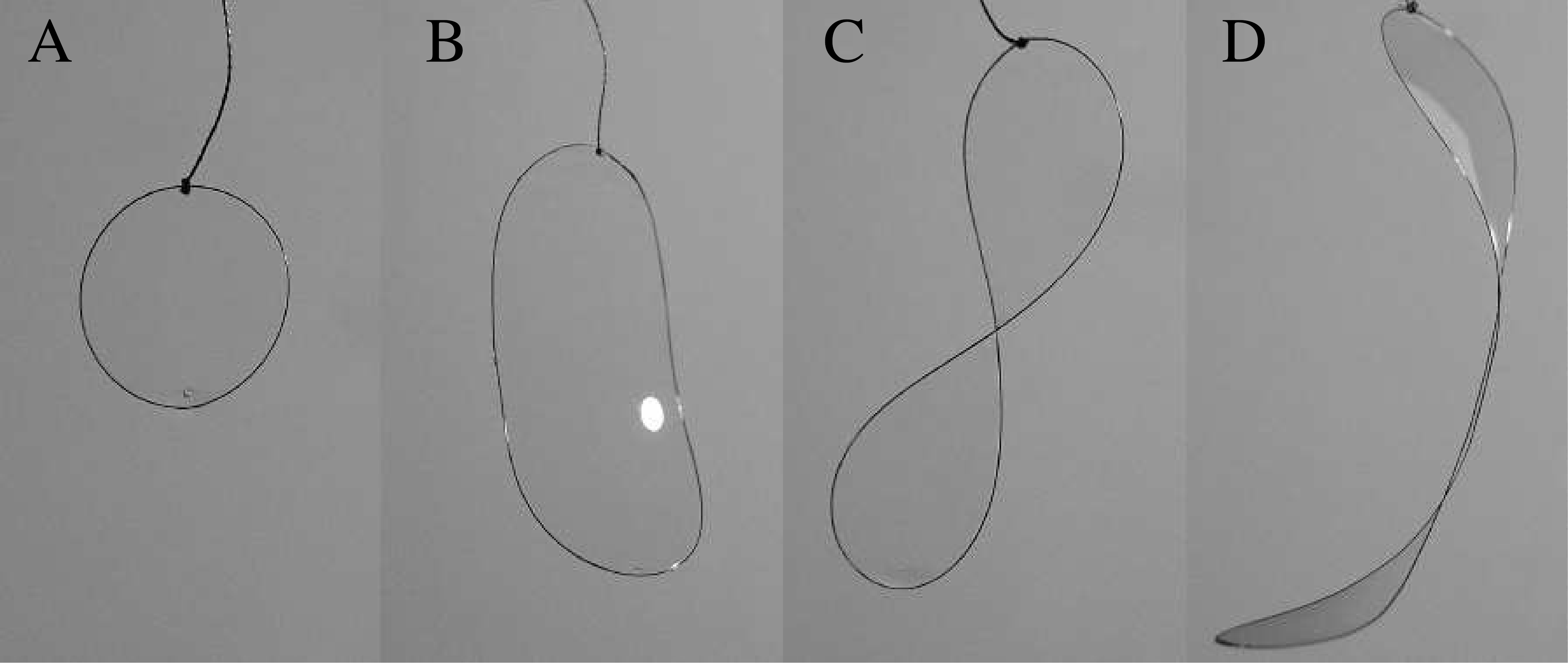}	
\caption{\label{fig:selection}A physical realization of the Euler-Plateau problem. (A)-(D) Examples of minimal surfaces bounded by an elastic line obtained by dipping (and removing) a loop of fishing line (0.3 mm diameter) in soapy water.  The interplay between the  forces due to surface tension of the film and the elastic forces arising at the boundary can be quantified via the dimensionless number $\hat{\sigma}=\sigma L^{3}/\alpha$ with $\sigma$ the surface tension, $L$ the length and $\alpha$ the bending rigidity of the boundary.  For small values of $\hat{\sigma}$, the soap film takes the form of (A) a planar disk, but upon increasing $\hat{\sigma}$ it buckles in the form of (B) a saddle, (C) a twisted figure eight and (D) a two headed racket-like structure. Reproduced from Ref. \cite{Giomi:2012}.}
\end{figure}

\section{\label{sec:ep}The Euler-Plateau Problem in Three Dimensions}

In Sec. \ref{sec:plane} we considered the problem of a soap film bounded by an elastic rod and constrained to lie on the plane. It is then very natural, and in fact tempting from a practical point of view, to lift the constraint of planarity and ask what would be the shape of a soap film bounded by an inextensible but flexible piece of wire that is free to move in three dimensions. This problem links two of the oldest problems in geometry and physics. Namely the problem of finding the shape of an {\em Elastica} subject to some external force and the Plateau problem of minimal surfaces. In geometry, the classical Plateau problem consists of finding the surface of least area that spans a given rigid boundary curve. A physical realization of the problem is obtained by dipping a stiff wire frame of some given shape in soapy water and then removing it. The original variational problem was formulated Lagrange \cite{Lagrange:1760} in the eighteenth century, however, the subject was lifted to a whole new level by Plateau in the mid-nineteenth century using a series of beautiful experiments which showed that a physical realization of these objects naturally arises through a consideration of soap films \cite{Plateau:1849}. Since then, the subject has inspired mathematicians \cite{Osserman:2002,Colding:2007,Morgan:2008}, scientists \cite{Thomas:1988,Kamien:2002,Penrose:1973}, engineers and artists alike~\footnote{Helaman Ferguson http://www.helasculpt.com}. 

Since experiments with soap films are often carried out by dipping closed wire frames into soap and then pulling them out, there is a natural generalization of this boundary-value problem that is suggested by the following question: what if the soap film is not bounded by a rigid contour, but has instead a soft boundary such as a flexible, inextensible wire? The resulting problem, that was named \emph{Euler-Plateau} problem in Ref. \cite{Giomi:2012}, provides a fascinating and highly nontrivial generalization of the planar problem reviewed in Sec. \ref{sec:plane}. 

Fig. \ref{fig:selection}A-D, shows a gallery of shapes  obtained via a simple experiment of dipping a loop of fishing-line into a solution of water and dish soap, and pulling it out. The resulting shapes are quite varied: short loops span a planar disk (Fig. \ref{fig:selection}A), intermediate loop lengths cause the film to spontaneously twist out of the plane (\ref{fig:selection}B) and eventually form a planar figure-eight shape (Fig. \ref{fig:selection}C). At the center of the eight (where the boundary crosses) the surface normal rotates by 180$^{\circ}$, and the spanning minimal surface is helicoid-like. Increasing the length even further leads to a a two-headed racket-like structure (Fig. \ref{fig:selection}D).

\subsection{Shape equation}

As in Sec. \ref{sec:plane} we start from the simple energy given in Eq. \eqref{eq:energy}, where $M$ is now a mapping of a two-dimensional disc in $\mathbb{R}^{3}$ and the boundary $\partial M$ is a space curve. For simplicity, we assume that the region of contact between the film and the bounding elastic rod is free to slide, so that the film can apply normal forces, but no torques. From this assumptions it follows that, if the filament is initially untwisted, it will remain untwisted even when buckled and bent out of the plane, although it can develop geometric torsion. To derive the Euler-Lagrange equations for the functional \eqref{eq:energy} we use again the Frenet frame introduced in Sec. \ref{sec:notation} and consider a small virtual displacement of the boundary: $\bm{r}\rightarrow\bm{r}+\xi\bm{n}+\eta\bm{b}$, where $\eta$ is a displacement along the binormal direction and vanishes identically if the film is constrained to lie on the plane. Calculating the linear variation yields \cite{Giomi:2012}:
\begin{multline}\label{eq:boundary-variation}
\delta \oint_{\partial M}ds\,(\alpha\kappa^{2}+\beta) 
= \oint_{\partial M} ds\,[\alpha(2\kappa''+\kappa^{3}-2\tau^{2}\kappa)-\beta\kappa]\xi\\
+ \alpha\oint_{\partial M}ds\,(4\kappa'\tau+2\kappa\tau')\eta\,.
\end{multline}
The linear variation of the area bounded by the curve can be expressed in a standard coordinate basis $\bm{g}_{i}=\partial_{i}\bm{R}$ (with $\bm{R}$ position vector of the film and $\partial_{i}$ partial derivative with respect to the $i$-th coordinate) as \cite{Lenz:2000}
\begin{align}
\delta \int_{M} dA 
&= \int_{M} dA\,(\nabla\cdot\bm{u}-2Hw) \nonumber \\
&= - \oint_{\partial M} ds\,\bm{T}\cdot\bm{u}-2\int_{M}dA\,H w\,.\label{eq:bulk-variation}
\end{align}
Here $\bm{u}$ and $w$ are respectively the displacement along the tangent plane of the film (expressed in the basis $\bm{g}_{i}$) and its normal direction $\bm{N}$, $H$ is the mean curvature of the film, $\bm{T}=\bm{N}\times\bm{t}$ is the tangent-normal vector of the Darboux frame defined in Sec. \ref{sec:notation} and the last equality comes from the divergence theorem on a surface. Compatibility demands that the bulk and boundary variations must be consistent with each other. Thus, using Eq. \eqref{eq:darboux-frame} to express $\bm{T}$ in the Frenet frame, after simple algebraic manipulations we obtain:
\begin{equation}\label{eq:compatibility}
\bm{T}\cdot\bm{u} = \eta\cos\vartheta\,\bm{b}-\xi\sin\vartheta\,\bm{n}\;,	
\end{equation}
where $\vartheta$ is the angle between the surface normal to the film and the normal to the boundary curve and will be herein referred to as {\em contact angle}. Enforcing the condition that the energy variations \eqref{eq:boundary-variation} and \eqref{eq:bulk-variation} vanish while satisfying the compatibility condition \eqref{eq:compatibility} yields:
\begin{subequations}\label{eq:euler-plateau}
\begin{align}
&\kappa''+\frac{1}{2}\kappa^{3}-\left(\tau^{2}+\frac{\beta}{2\alpha}\right)\kappa-\frac{\sigma}{2\alpha}\,\sin\vartheta = 0\,, \\[5pt]	
&2\kappa'\tau+\kappa\tau'+\frac{\sigma}{2\alpha}\,\cos\vartheta = 0\,, \\[10pt]
&H=0.
\end{align}
\end{subequations}
Alternatively, one can write Eqs. (\ref{eq:euler-plateau}a,b) in a form that explicitly contains the geodesic and normal curvature of the boundary by expressing $\cos\vartheta=\kappa_{n}/\kappa$ and $\sin\vartheta=\kappa_{g}/\kappa$.

Unlike the planar case, the Euler-Lagrange equations \eqref{eq:euler-plateau} are very difficult and perhaps impossible to solve analytically for the general case. It is possible nonetheless to infer a number of properties without  knowledge of the exact solution. Multiplying Eq. (\ref{eq:euler-plateau}b) by $\kappa$ allows us to recast the resulting expression as $(\kappa^{2}\tau)'=-(\alpha/2\sigma)\kappa_{n}$, which integrated along the length of the closed boundary yields the following integral formula for the normal curvature $\kappa_{n}$:
\begin{equation}\label{eq:total-normal}
\oint_{\partial M} ds\,\kappa_{n} = 0\,,
\end{equation}
valid for all values of the physical parameters. Furthermore, it is possible to consider a special class of solutions of Eqs. \eqref{eq:euler-plateau} for which the contact angle $\vartheta$ is constant along the curve. Then Eq. \eqref{eq:total-normal} can be used to prove that $\vartheta$ must then be $90^{\circ}$. To understand this latter statement we first need to recall that the normal vector $\bm{n}$, thus the contact angle $\vartheta$, is undefined at inflection points (i.e. where $\kappa=0$, see Sec. \ref{sec:notation}). In particular, if $\vartheta$ is constant, $\kappa$ must be everywhere positive. Then, writing $\kappa_{n}=\kappa\cos\vartheta$ in Eq. \eqref{eq:total-normal} yields:
\begin{equation}
\cos\vartheta\oint_{\partial M} ds\,\kappa = 0\,, 
\end{equation}
but since $\kappa$ is strictly positive, this last condition is possible if and only if $\cos\vartheta=0$, thus $\vartheta=\pm \pi/2$: i.e. $\kappa_{n}=0$ and the surface normal must be perpendicular to the curve normal everywhere if $\vartheta$ is to be constant. This imposes strong constraints on the nature of the solution in this case as we will now argue.

Curves having zero normal curvature everywhere are called {\em asymptotic curves} and their tangent vectors {\em asymptotic directions}. Every curve that lies on the plane is clearly asymptotic. On a minimal surface, on the other hand, there are two asymptotic curves passing through each point where the Gaussian curvature $K$ is non-zero and these curves always intersect at 90$^{\circ}$. If the minimal surface contains flat points (i.e. isolated points where $K=0$), these are traversed by $\mathcal{N}>2$ asymptotic curves crossing at $\pi/\mathcal{N}$ and forming, in their neighborhood, $n$ valleys separated by ridges \cite{Koch:1990}. These observations suggest that the only simply connected minimal disk bounded by a closed asymptotic curve lies in the plane;  otherwise the asymptotic directions in the disk would form a vector field tangent to the boundary. In a simply connected disk this vector must then enclose a singularity or vortex, in the neighborhood of which the structure of the asymptotic directions would disagree with the previous classification. Hence, we conclude that the only solution having constant contact angle is given by the flat disk.

\subsection{Numerical simulations of the soap film shapes}

To explore the variety of possible shapes resulting from the solution of Eqs. \eqref{eq:euler-plateau} and the transitions between them, Mahadevan and the author \cite{Giomi:2012} minimized a discrete analog of the total energy \eqref{eq:energy}. The soap film is approximated as simplicial complex consisting of an unstructured triangular mesh. The internal edges of the triangles are treated as elastic springs of zero rest-length so that the total energy of the mesh is given by:
\begin{equation}\label{eq:discrete-energy}
E = \alpha\sum_{v \in \partial M} \langle s_{v} \rangle \kappa_{v}^{2} + k \sum_{e \in M } |e|^{2}\,.	
\end{equation}	
The first sum runs over the boundary vertices and $\langle s_{v} \rangle = (s_{v}+s_{v-1})/2$ is the average of the length of the two edges meeting at $v$. The curvature of the boundary is calculated as $\kappa_{v}=|\bm{t}_{v}-\bm{t}_{v-1}|/\langle s_{v} \rangle$ with $\bm{t}_{v}$ and $\bm{t}_{v-1}$ the tangent vectors at $v$. The second sum in Eq. \eqref{eq:discrete-energy} runs over all internal edges. If the triangles are equilateral, this yields a discrete approximation for the soap-film energy with the spring stiffness proportional to the surface tension, i.e. $\sigma \approx 4k\sqrt{3}/(2-B/E)$, where $B/E$ is the ratio between the number of boundary edges $B$ and the total number of edges $E$ of the triangular mesh. {The choice of minimizing the squared length of the edges, instead of the area of the triangles, is motived by numerical stability. Replacing the spring energy in \eqref{eq:discrete-energy} with the sum of the area of the triangles, has the effect of shortening the range of interaction between the boundary and the interior to the single strip of triangles at the boundary. Most standard local optimization algorithms would then attempt to reduce the area of these triangles to zero, thus suppressing the interaction between the boundary and the interior.  In the continuum limit, the energy \eqref{eq:discrete-energy} approaches that given by \eqref{eq:energy} for  original problem, and  the sequence of shapes obtained with this method is in excellent agreement with our experimental observations shown in Fig. \ref{fig:selection}.

Numerical minimization of Eq. \eqref{eq:discrete-energy} using a conjugate gradient method leads to a variety of shapes depending on the values of $\alpha$, $k$, the boundary length $L$ as well as the initial shape of the domain. An interactive gallery of the shapes obtained from the numerical simulations is available on-line~\footnote{\url{http://www.seas.harvard.edu/softmat/Euler-Plateau-problem/}}. A first set of simulations were run by starting with an unstructured triangular mesh of 474 vertices bounded by an {\em elongated} hexagonal polygon; the mesh points are started out randomly displaced transverse to the plane. As the spring constant $k$ was increased, corresponding to an increase in the  surface tension, the system moved through a series of transitions shown in Fig. \ref{fig:twisting}. Beyond a critical value of the surface tension (relative to the bending stiffness, as discussed earlier), the planar disk  buckled into a two-fold  mode consistent with the linear stability analysis reported in Sec. \ref{sec:analytic-solution}. For $kL^{3}/\alpha\approx 643$,  the planar elliptical shape transitions to a twisted non-planar saddle-like shape, as experimentally observed. This is accompanied by an increase in the normal curvature of the boundary and there is  a progressive twisting of the central ``waist'' of the surface. A suitable quantity to characterize the non-planarity of the boundary is obtained by integrating its absolute normal curvature:
\begin{equation}\label{eq:order-parameter}
\langle |\kappa_{n}| \rangle = \oint_{\partial M} ds\,|\kappa_{n}|\,.
\end{equation}
A plot of $\langle |\kappa_{n}| \rangle$ for a wide range of $kL^{3}/\alpha$ values is shown in Fig. \ref{fig:twisting}. A further increase in $k$ leads to an increase in the twist  until the surface normal undergoes a full 180$^{\circ}$ rotation across the waist for $kL^{3}/\alpha\approx 740$.  The surface then reverts to a planar configuration in the shape of the number eight (see Fig. \ref{fig:twisting}E). 

\begin{figure*}[t]	
\centering
\includegraphics[width=0.8\textwidth]{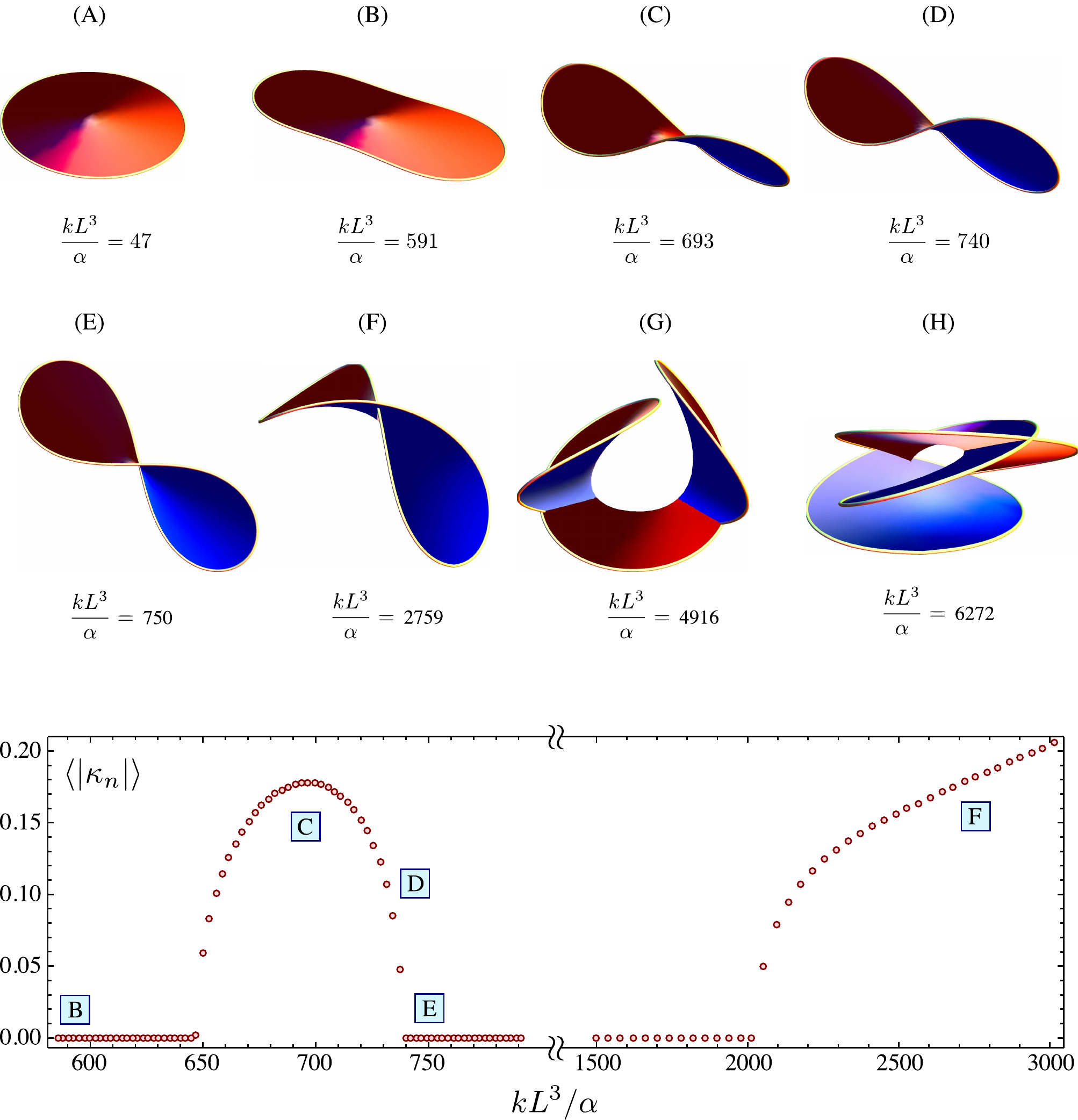}	
\caption{\label{fig:twisting}Some solutions of the Euler-Plateau problem and a bifurcation diagram. Shapes obtained by minimizing the discrete approximation \eqref{eq:discrete-energy} of the continuum energy of a minimal surface bounded by an elastic line \eqref{eq:energy} as a function of the dimensionless number $kL^{3}/\alpha$. For small values of the parameter, the flat circular disk of radius $L/2\pi$ has the lowest energy. Upon increasing $kL^{3}/\alpha$ the disk first buckles into flat elliptical shape (see text). For $kL^{3}/\alpha>643$, this elliptical disk  deforms into a three-dimensional saddle-like shape, becoming increasingly twisted  with an increase in  $kL^{3}/\alpha>643$ until the surface eventually becomes flat again, but now forming a fully twisted eight-shaped conformation. This twisted flat shape is a local energetic minimum for $740<kL^{3}/\alpha<2000$. For still larger $kL^{3}/\alpha$ the twisted surface self-intersects at the waist of the figure-eight and the two lateral lobes bend toward each other, while for very large $kL^{3}/\alpha$ the surface exhibits several self-intersections leading to the complex structure (H). Below the shapes we show the bifurcation diagram for the system, characterizing the shape using the absolute normal curvature of the boundary integrated along its length (a measure of the amplitude of the instability) as a function of the bifurcation parameter $kL^{3}/\alpha$, along with the location of the transitions described above. Reproduced from Ref. \cite{Giomi:2012}.}
\end{figure*}

Once the surface reaches the planar figure-eight conformation an increase in the surface tension does not produce further conformational changes until $kL^{3}/\alpha\approx 2000$. Beyond this value, the surface again becomes non-planar, self-intersects at the pinch of the eight, and the two lateral lobes start bending toward each other. The curvature of the lobes increases with increasing $k$ until  they intersect to produce the complex shape shown Fig. \ref{fig:twisting}H. The transition to this latter surface is analogous to the transition from the figure-eight to the two headed racket-like structure observed in the experimental realization of the problem. However, in the experimental system, self-intersection cannot occur and adhesion favors the formation of a line of contact between the two lobes (see Fig. \ref{fig:selection}D). 

Not surprisingly, given the nonlinear nature of the governing equations \eqref{eq:euler-plateau}, the solutions are not expected to be unique. Numerically, one finds that the final equilibrium state strongly depends on the initial configuration of the system.  If the initial configuration is {\em not elongated}, as in say a triangular mesh bounded by a hexagon, which is closer to being circularly symmetric, other possible stable configurations are found. As in the previous case, for small values of the surface tension, the system rapidly relaxes into a flat circular disk. However, for $kL^{3}/\alpha\approx 855$ the system transitions to a saddle-like configuration that is the classical Enneper minimal surface. Upon increasing the surface tension still further, the curvature of the boundary and film becomes larger still and the surfaces self-intersect leading to the beautiful structure shown in Fig. \ref{fig:enneper}D.

\begin{figure*}[t]	
\centering
\includegraphics[width=0.8\textwidth]{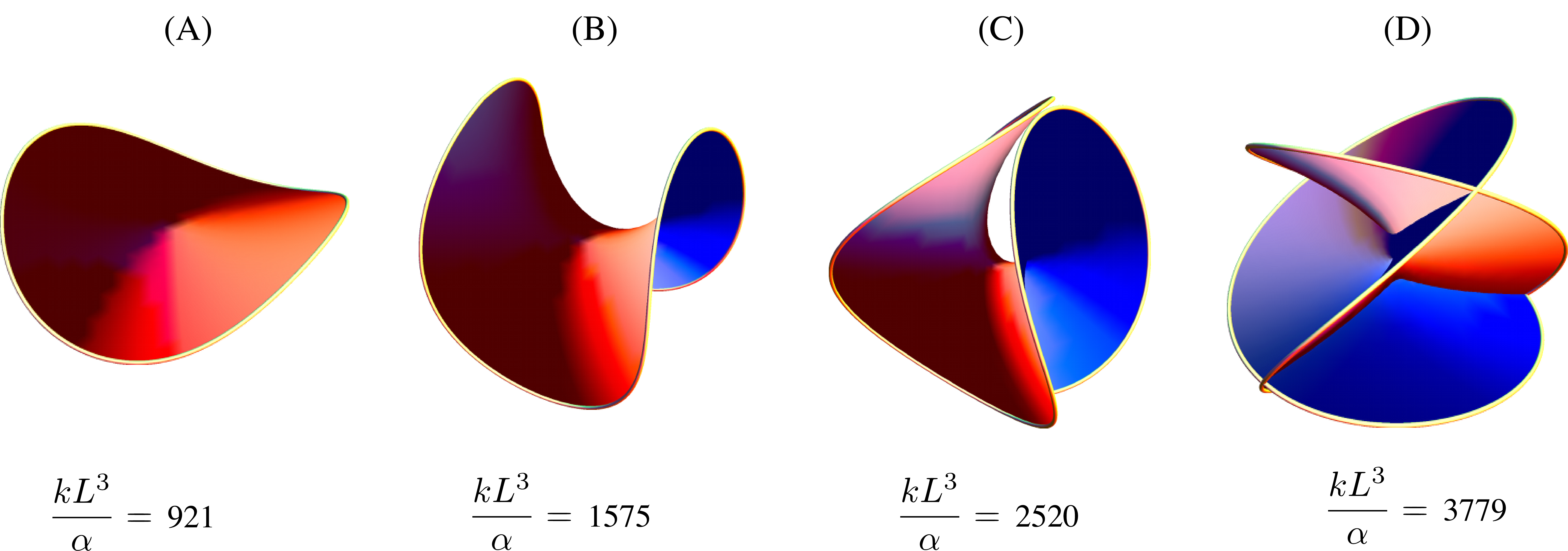}	
\caption{\label{fig:enneper}Further equilibrium solutions of the Euler-Plateau problem. An initial configuration of a triangular mesh bounded by a hexagonal perimeter gives the classic Enneper minimal surface rather than the figure-eight shape shown in Fig. \ref{fig:twisting}. Reproduced from Ref. \cite{Giomi:2012}.}
\end{figure*}
 
\subsection{Asymptotic analysis}

The numerical data show that the transition from the planar two-fold symmetric shape to the non-planar twisted eight is consistent with a supercritical pitchfork bifurcation, with $\langle |\kappa_{n}|\rangle \sim (\hat{\sigma}-\hat{\sigma}_{\rm t})^{1/2}$, where the angular brackets stand for an average along the boundary curve. Indeed Fig. \ref{fig:bifurcation} shows a best fit of the numerical data at the onset of the transition, the exponent obtained from the fit is $0.489 \pm 0.005$. 

To understand the origin of the pitchfork bifurcation, it is helpful to introduce a specific approximate representation of the twisted soap film in terms of the following one-parameter family of surfaces: 
\begin{subequations}
\begin{align}\label{eq:eight}
x &= r\,(1+t^{2})\cos\phi\,,\\[5pt]
y &= r\,(1-t^{2})\sin\phi\,,\\[5pt]
z &= t\left(r^{2}/R\right)\sin 2\phi\,, 	
\end{align}
\end{subequations}
where $0\le 0 \le R$ and $0 \le \phi < 2\pi$ are the usual plane-polar coordinates. Here $-1 \le t \le 1$ is a parameter that characterizes the family of shapes. {Eq. \eqref{eq:eight} is the simplest parameterization of a twisted saddle with disk topology and whose degree of non-planarity is controlled by the parameter $t$. For $t=0$, the surface given by (\ref{eq:eight}) is a disk of radius $R$; as $t$ becomes positive (negative), the surface becomes a right-handed (left-handed) twisted figure-eight, and for $t=\pm 1$ the surface reduces to a flat twisted figure-eight bounded by a lemniscate of Gerono. In the following we will exploit the topological equivalence between this family of surfaces and the actual twisted soap film to study the global properties of the surface at the onset of the transition, when $|t|\ll 1$.

Lengthy, but straightforward calculations give an expression for the curvatures of the boundary up the fourth order in $t$:
\begin{gather*}
\kappa_{n} \approx-\frac{2t}{R}\sin 2\phi+\frac{2t^{3}}{R}(3\sin 2\phi-\sin 4\phi+\sin 6\phi)\;,\\[5pt]
\kappa_{g} \approx \frac{1}{R}+\frac{t^{2}}{R}(1+3\cos 2\phi-5\cos 4\phi)\;	
\end{gather*} 
Similarly, the arc-length is given by:
\begin{equation}
ds^{2} = R^{2}[(1+t^{2})^{2}-2t^{2}(\cos 2\phi-\cos 4\phi)]\;. 	
\end{equation}
From this one can calculate an approximated expression for the total energy at the onset of twisting (i.e. for $t\sim 0$). Namely:
\[
E \approx E_{0}+\frac{\pi\alpha}{R}\left[t^{2}\left(10+\frac{\sigma R^{3}}{\alpha}\right)-t^{4}\left(9+\frac{5}{3}\frac{\sigma R^{3}}{\alpha}\right)\right]\;,
\]
where $E_{0}$ is again the energy of the circular configuration. Finally, the condition of inextensibility for the boundary curve is given by:
\begin{equation}\label{eq:approximated-length}
L = \oint_{\partial M} ds \approx 2\pi R \left[ 1+t^{2}\left(1-\tfrac{1}{2}t^{2}\right)\right]\;.
\end{equation}
Using equation Eq. \eqref{eq:approximated-length} to eliminate $R$ and then taking the derivative of the approximate energy with respect to $t$ yields the following equation of equilibrium for the amplitude parameter $t$ that characterizes the shape of the boundary:
\begin{equation}\label{eq:normal-form}
t(96\pi^{3}-\hat{\sigma})+\tfrac{2}{3}\,\hat{\sigma} t^{3} = 0\;,
\end{equation}
which is the normal form of a supercritical pitchfork bifurcation and implies:
\begin{equation}\label{eq:bifurcation}
\langle |\kappa_{n} |\rangle \sim (\hat{\sigma}-\hat{\sigma}_{\rm t})^{1/2}\;,\qquad
\hat{\sigma}_{\rm t} = 96\pi^{3}\;,
\end{equation}
having used that $\langle |\kappa_{n}| \rangle \sim t$. Analogously one can calculate the behavior of the Gaussian curvature of the bounded minimal surface at the onset of twisting. Using the Gauss-Bonnet \cite{DoCarmo:1976} one has:
\begin{equation}
\int_{M} dA\,K = 2\pi-\oint_{\partial M} ds\,\kappa_{g} \approx 4\pi t^{2}\;,
\end{equation}
from where, the integrated Gaussian curvature $\langle K \rangle=\int dA\,K$ is given by:
\begin{equation}
\langle K \rangle \sim (\hat{\sigma}-\hat{\sigma}_{\rm t})\;,
\end{equation}
in excellent agreement with the numerical simulations shown in Fig. \ref{fig:bifurcation}.

\begin{figure}[t]
\centering
\includegraphics[width=0.9\columnwidth]{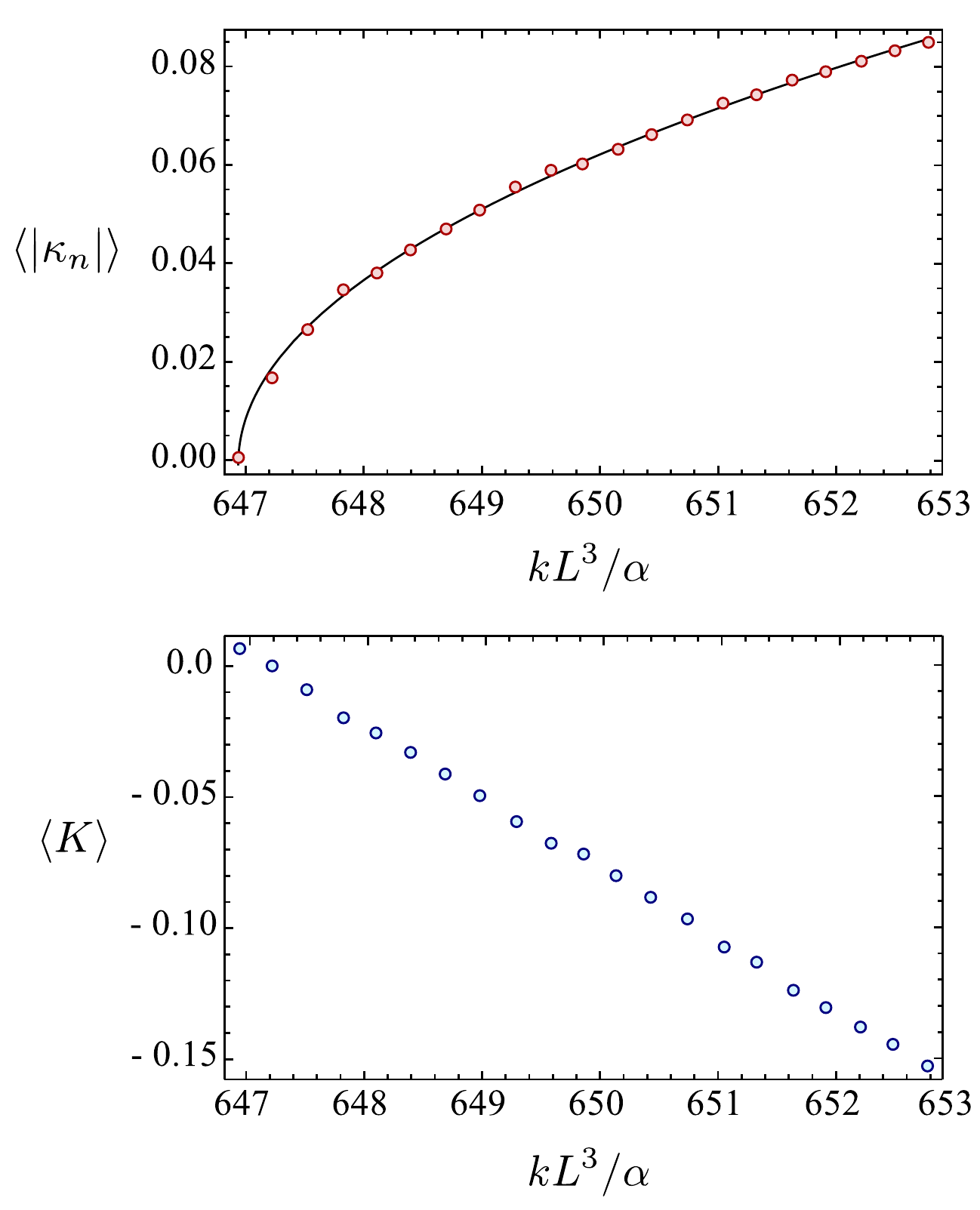}	
\caption{\label{fig:bifurcation}Characterizing the onset of twist and non-planarity. The left panel shows the average absolute normal curvature of the boundary $\langle |\kappa_{n}| \rangle$  as a function of $kL^{3}/\alpha$ in the neighborhood of the onset of the transition to non-planarity. The solid line on the left figure is obtained from a best fit of the data. The exponent resulting from the fit is $0.489 \pm 0.005$, and is consistent with the instability being of a supercritical pitchfork type. The right panel shows the average Gaussian curvature of the surface $\langle K \rangle$ as a function of $kL^{3}/\alpha$ in the neighborhood of the onset of the transition to non-planarity. Reproduced from Ref. \cite{Giomi:2012}.}
\end{figure}

\subsection{Final remarks}

The Euler-Plateau problem, reviewed in this section, is a fascinating example of a free boundary value problem obtained from the natural marriage of two classic problems of geometry and physics. While a minimal physical realization of the Euler-Plateau problem is a kitchen-sink experiment, its ramifications are likely to go far beyond this specific manifestation, just as the mechanics of soap films and elastic filaments have been relevant for the study of matter, not just at the every-day scale, but also for systems that range from molecules \cite{Thomas:1988} to black holes \cite{Penrose:1973}.  However, unlike its planar version reviewed in Sec. \ref{sec:plane}, the three-dimensional problem appears as a formidable mathematical challenge and the open questions outnumber the answers by far. Some immediate mathematical questions include proofs of existence, regularity of solutions, as well as measures of the non-uniqueness of solutions, particularly for large values of the only parameter in the problem $\hat{\sigma}$ or its discrete analogue $kL^{3}/\alpha$. 

\begin{figure}[t]
\centering
\includegraphics[width=1\columnwidth]{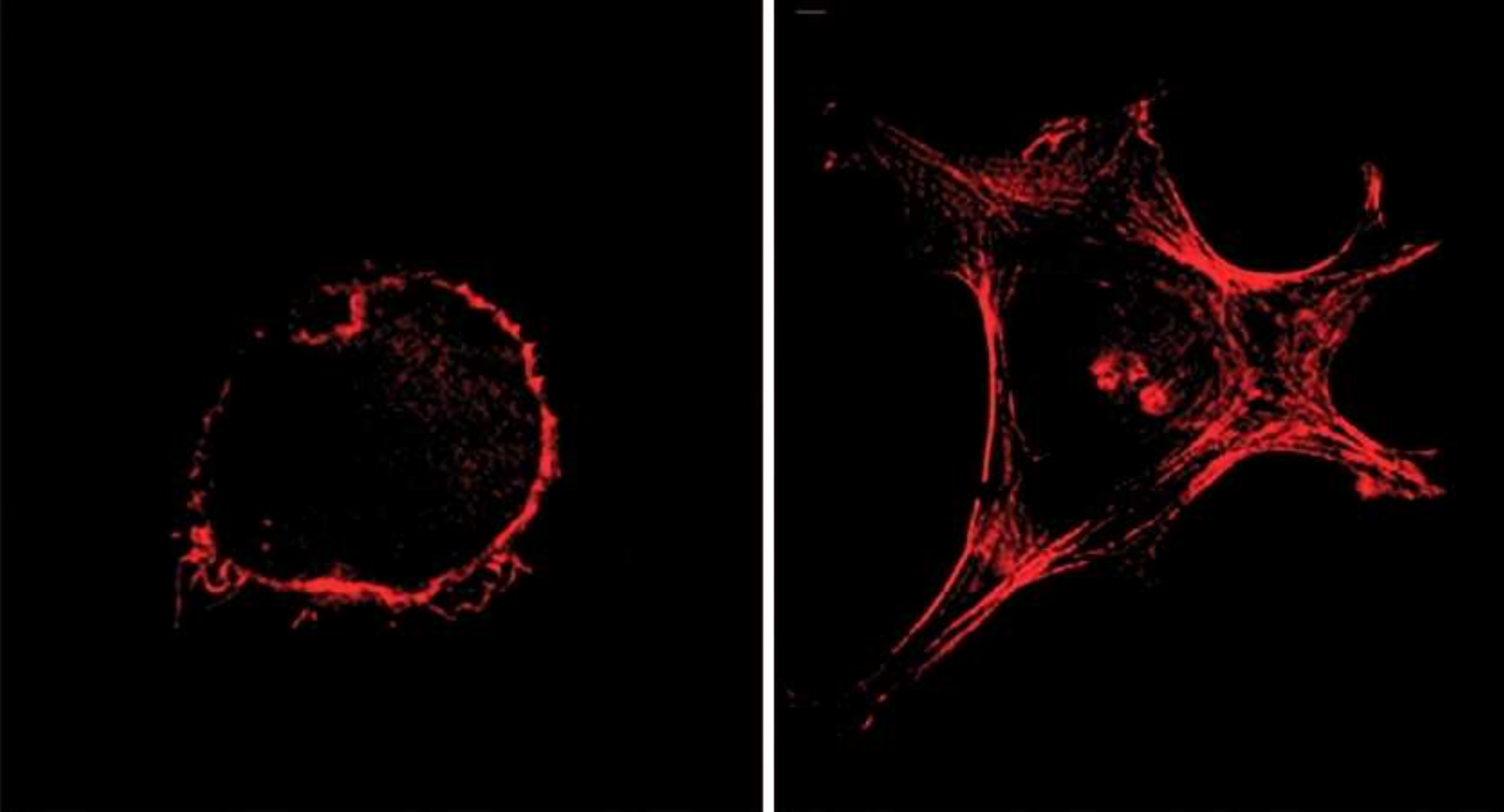}	
\caption{\label{fig:cell}Fluorescently labeled actin cytoskeleton in fibroblasts adhering on a soft (left) and stiff (right) substrate. On the soft substrate the cell has a modest spread area and a rounded morphology, while on the stiffer substrate the cell spreads more and its geometry is concave and branched. Reproduced with permission from Ref. \cite{Discher:2005}.}
\end{figure}

\section{\label{sec:cell}Application to Cell Mechanics}

As we pointed out in the previous sections, determining the optimal shape of a soap film bounded by an elastic element is not just an intriguing mechanical puzzle inspired by the observation of the mundane, but an important prototype problem for various physical systems whose geometry is softly constrained. In this section we will consider an important application of these methods to the case of contractile cells adhering to compliant substrates.

The propagation of mechanical forces generated during cell-matrix adhesion is now recognized as one of the fundamental mechanism beyond a variety of cellular processes. By mean of adhesive contacts with the substrate, a cell is able to sense mechanical cues from the environment, transmit them to its interior and transduce them into chemical signals that in turn activate a spectrum of biological responses \cite{Discher:2005,Janmey:2007,Asano:2009}. For instance cells adhering to softer substrates spread less and prefer to have well rounded morphologies, while they are more likely to exhibit branched patterns on stiffer substrates with greater spread area~\cite{Yeung:2005,Chopra:2011} (Fig. \ref{fig:cell}). Experiments on micro-patterned adhesive islands revealed that cell proliferation and spreading sensitively depend on adhesion geometry~\cite{Chen:1997}. Perhaps the most impressive effect of the cell-substrate interaction is its ability to determine the fate of some type of stem cells that, adapting themselves to the substrate stiffness, might becomes either neurons or bone cells \cite{Engler:2006}.

\subsection{The contractile film model for adherent cells}

Fully spread cells adhering on a solid substrate generally exhibit a flat morphology with the nucleus forming bump in the center. Adhesion with the substrate predominately occurs along the cell boundary in a discrete set of adhesion sites or continuously. This suggests that insight about the geometry and the mechanics of cell adhesion can be gained, in first approximation, by treating the cell as a two-dimensional film that is in contact with the substrate through points localized along its contour. At time scales when the cell is fully spread, the forces generated by the actomyosin cytoskeleton are mainly contractile and this gives rise to an {\em effective tension} that is transmitted to the substrate thorough the focal adhesions \cite{Barziv:1999}. The actin cortex localized at the cell periphery naturally resists to the inward contraction by opposing an elastic force. The overall effect of actomyosin contractily, cortex elasticity and adhesion on the shape of the cell was investigated in Ref. \cite{Banerjee:2013}, by mean of a simple and yet very rich mechanical model inspired by the planar film problem reviewed in Sec. \ref{sec:plane}: the Contractile Film Model. In this model, the shape of the cell-substrate contact line is parametrized by closed plane curve and the total mechanical energy of the system is approximated, on the basis of symmetry arguments, in the following form:
\begin{equation}\label{eq:cfm}
E = \sigma\int_{M}dA\,+\oint_{\partial M}ds\,(\alpha\kappa^{2}+\beta+k_{s}\rho\,|\bm{r}-\bm{r}_{0}|^{2})\,.
\end{equation}
Here $\sigma$ represents the {\em effective} surface tension in the cell due to cytoskeletal contractility, while the last term in Eq.~\eqref{eq:cfm} represents the strain energy induced by the cell on a substrate of stiffness $k_{s}$ through focal adhesions localized at the cell edge~\cite{Wozniak:2004} with density $\rho(s)$, so that the total number of adhesions is $N_{A}=\oint ds \rho$. The reference configuration $\bm{r}_0(s)$ describes the shape attained by the cell boundary on a rigid substrate, and is a constraint set by the adhesion geometry. It is important to stress that all the physical constants in Eq. \eqref{eq:cfm} are {\em effective} as, in addition to the normal mechanical properties of the passive elements (the cell membrane, the liquid component of the cytosol etc.), they account for the additional forces due to the actomyosin activity. 

A class of theoretical models by Bischofs {\em et al}. \cite{Bischofs:2008,Bischofs:2009} have been constructed around the competition of bulk and peripheral contractility and ignored the bending elasticity of the actin cortex (i.e. $\alpha=0$). In analogy with the Laplace law of capillarity, the steady state cell contour is then described by concave circular arcs of radius $\beta/\sigma$ connecting adhesion sites. The model introduced in Ref. \cite{Banerjee:2013} and reviewed here, instead, focuses on the opposite limit and considers the regime in which the force balance is dominated by the competition between cortex elasticity and bulk contractility, while the effect of peripheral contractility is negligible (i.e. $\beta=0$). In this scenario, the curvature is generally non-uniform, especially in the neighborhood of adhesion sites. As we will see in Sec. \ref{sec:discrete}, incorporating bending elasticity leads to an extremely rich polymorphism and allows for a transition from purely convex to purely concave cell shape reminiscent of that observed in experiments on cardiac myocytes~\cite{Chopra:2011}. From a purely mechanical point of view, the Contractile Film Model defined by Eq. \eqref{eq:cfm} is equivalent to a fluid film bounded by a planar {\em Elastica} and attached to a continuous or discrete set of links to a compliant substrate. Unlike the problem reviewed in Sec. \ref{sec:plane}, however, Eq. \eqref{eq:cfm} does not involve any hard constraint on the perimeter of the cell, which is then only softly constraint by the adhesion with the substrate. 

\subsection{Continuous adhesion}
\begin{figure}[t]
\centering
\includegraphics[width=0.7\columnwidth]{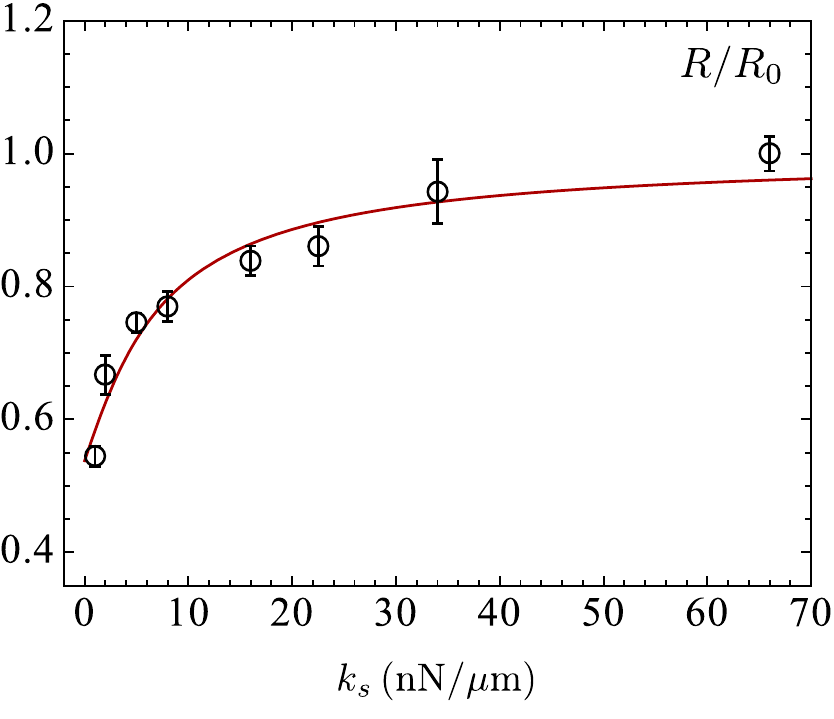}
\caption{\label{fig:spreading} Relative cell size $R/R_0$ as a function of substrate stiffness $k_s$ (solid black circles) for smooth muscle cells, 4 hours after plating on continuous elastic gels~\cite{Engler:2004}. Cell radius is estimated from the projected cell area reported in Ref. \cite{Engler:2004} as $R=\sqrt{\text{area}}/\pi$. Substrate stiffness $k_s$ is determined from substrate Young's modulus $E_s$ as : $k_s=a E_s$, where $a$ is the characteristic focal adhesion size, with $a\sim 1\ \mu m$. Solid (red) line represents the solution to Eq.~\eqref{eq:radius} with with $\sigma= 1.05$ nN/$\mu$m and $\alpha/R_0^3=0.16$ nN/$\mu$m. Reproduced form Ref. \cite{Banerjee:2013}.}
\end{figure}
\begin{figure}[t]
\centering
\includegraphics[width=\columnwidth]{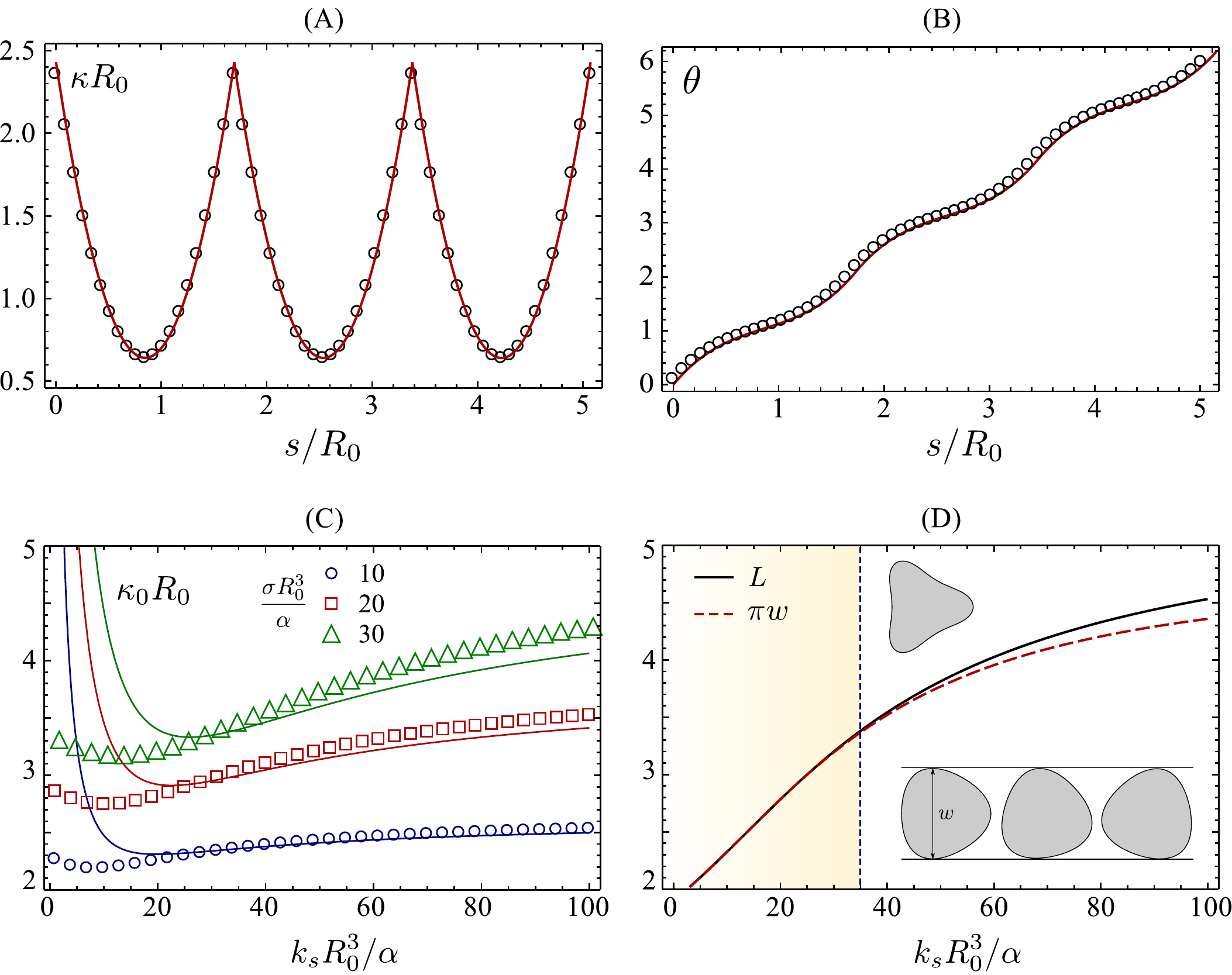}
\caption{\label{fig:geometry}Cell anchored onto three pointwise adhesions located at the vertices of an equilateral triangle. The curvature (A) and the tangent angle (B) as function of arc-length for $\sigma R_{0}^{3}/\alpha=10$, $k_{s}R_{0}^{3}=50$ and $N_{A}=3$. The circles are obtained from a numerical minimization of a discrete version of the energy \eqref{eq:energy}, while the solid lines corresponds to our analytical approximation. (C) The total cell length $\mathcal{L}$ as a function of adhesion stiffness. For small stiffnesses the cell boundary form a curve of constant width (lower inset) and $\mathcal{L}=\pi w$, with $w$ the width of the curve. This property breaks down for larger stiffnesses when inflection points develops (upper inset). (D) The curvature $\kappa_{0}$ at the adhesion points as a function of the substrate stiffness for various contractility values. The points are obtained from numerical simulations while the solid lines correspond to our analytical approximation. Reproduced from Ref. \cite{Banerjee:2013}.}
\end{figure}
\begin{figure*}[t]
\centering
\includegraphics[width=0.9\textwidth]{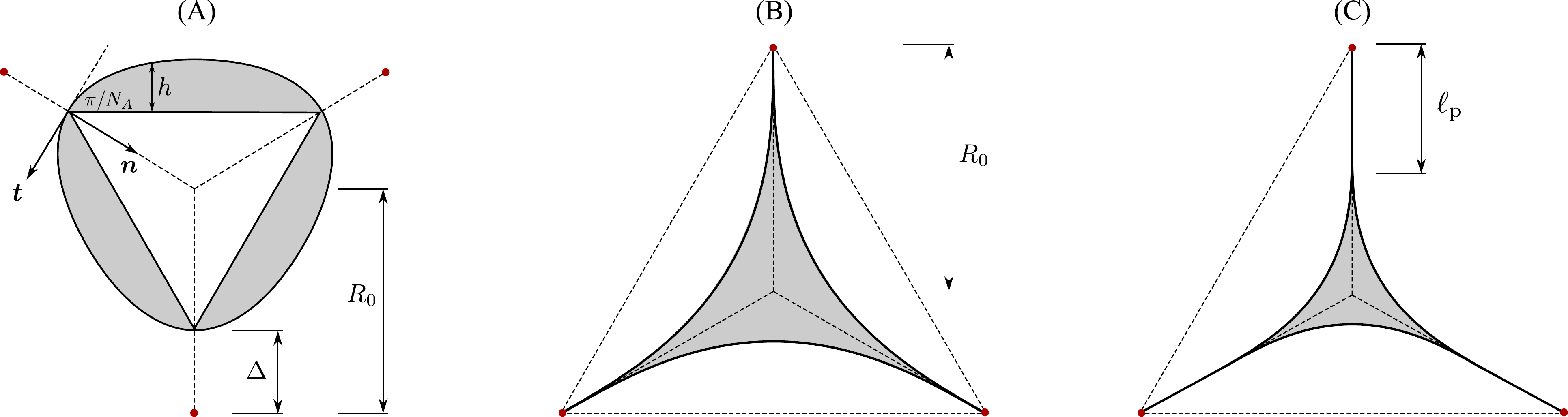}
\caption{\label{fig:schematic}(a) $\sigma < \sigma_{c1}$, cell contour is everywhere convex with constant width. (b) $\sigma=\sigma_p$, cell contour is purely concave with cusps at adhesion points and without protrusions. (c) $\sigma>\sigma_{c2}$, cusps are connected to the substrate by means of a protrusion of length $\ell$.}
\end{figure*}
If the periphery of the cell forms contact with a single continuous adhesion site, so that $\rho=1/L$ with $L$ the perimeter of the cell, the energy minimization problem admits a simple solution. In the presence of a uniform and isotropic substrate, we can assume the reference configuration to be a circle of radius $R_{0}$ so that a natural minimizer of the energy \eqref{eq:cfm} would be a circle or radius $R$. Thus, setting $\beta=0$, $\kappa=R^{-1}$ and $\rho^{-1}=2\pi R$ and minimizing Eq. \eqref{eq:cfm} yields the following cubic equation:
\begin{equation}\label{eq:radius}
(k_{s}+\pi\sigma)R^{3}-k_{s}R_{0}R^{2}-\pi\alpha = 0\;,
\end{equation}
The equation contains two length scales, $R_{0}$ and $\xi=(\alpha/\sigma)^{1/3}$, and a dimensionless control parameter $k_{s}/\sigma$ expressing the relative amount of adhesion and contraction. For very soft anchoring $k_{s}\ll \sigma$ and Eq.~\eqref{eq:radius} admits the solution $R=\xi$, whereas if the cell is rigidly pinned at adhesion sites, $k_{s}\gg \sigma$ and $R\rightarrow R_0$. For intermediate values of $k_{s}/\sigma$ the optimal radius $R$ interpolates between $\xi$ and $R_{0}$ and is an increasing function of the substrate stiffness $k_{s}$, in case $\xi<R_{0}$, or a decreasing function if $\xi>R_{0}$. For $\xi=R_{0}$, the lower and upper bound coincide, and the solution is $R=R_0$. In particular, the case $R_0>\xi$ reproduces the experimentally observed trend that cell projected area increases with increasing substrate stiffness before reaching a plateau at higher stiffnesses \cite{Yeung:2005,Chopra:2011}. 

Fig. \ref{fig:spreading} shows a fit to the measured projected areas of smooth muscle cells (SMCs) adhering to continuous elastic gels of varying substrate elastic modulus \cite{Engler:2004} with the solution of Eq. \ref{eq:radius}. The data for the spread area of SMCs are taken 4 hours after plating onto the substrate, when they retain rounded morphologies. The fitted value for surface tension $\sigma=1.05$ nN/$\mu$m comes to the same order of magnitude as reported for endothelial cells~\cite{Lemmon:2005,Bischofs:2009}, epithelial cells~\cite{Mertz:2012} and is consistent with the numerical estimate provided earlier. The fit also provides a value for the bending rigidity $\alpha=4.62 \times 10^{-16}$ Nm$^2$. The asymptotic behavior and various limits of the solution are well captured by the interpolation formula:
\begin{equation}
R\approx\frac{k_{s}R_{0}+3\pi\sigma\,\xi}{k_{s}+3\pi\sigma}
\end{equation}
indicating that larger surface tension, hence larger cell contractility $\sigma$ leads to lesser spread area, consistent with the experimental observation that myosin-II activity retards the spreading of cells~\cite{Wakatsuki:2003}. Standard stability analysis of this solution under a small periodic perturbation in the cell radius shows that the circular shape is always stable for any values of the parameters $\sigma$, $k_{s}$ and $R_{0}$.

\subsection{\label{sec:discrete}Discrete adhesions}

For cells adhering to discrete number of adhesion sites, one can show that the circular solution for the cell boundary is never stable and there is always a non-circular configuration with lower energy. For simplicity, we assume that $N_{A}$ adhesion sites are located at the vertices of a regular polygon of circumradius $R_{0}$, with density $\rho(s)=\sum_{n=0}^{N_{A}-1}\delta(s-i\ell)$,  and $\ell$ the distance between subsequent adhesions. Following the derivation given in Sec. \ref{sec:shape-equation} and taking into account that $\delta(\oint ds\,\rho)=0$, since the number $N_{A}$ of adhesions is assumed constant, the Euler-Lagrange equation can be obtained in the form: 
\begin{equation}\label{eq:euler-lagrange}
\alpha\left(2\kappa''+\kappa^{3}\right)- \sigma + 2 k_s\sum_{i=0}^{N_{A}-1}  \delta(s-i\ell)\left({\bm r}-{\bm r}_0\right)\cdot {\bm n}= 0\;.
\end{equation}
Due to the $N_{A}$-fold symmetry of the adhesion sites, adhesion springs stretch by an equal amount $\Delta$ in the direction of the normal vector: $(\bm{r}_{i}-\bm{r}_{i0})\cdot\bm{n}_{i}=\Delta$, $i=1,\,2\ldots N_{A}$. As a consequence of the localized adhesion forces, the curvature is non-analytical at the adhesion points. Integrating Eq. \eqref{eq:euler-lagrange} along an infinitesimal neighborhood of a generic adhesion point $i$, one finds:
\begin{equation}\label{eq:discontinuity}
\kappa'_{i}=-\frac{k_{s}}{2\alpha}\,\Delta\;.
\end{equation}
The local curvature of the segment lying between adhesion points is on the other hand determined by Eq. \eqref{eq:shape-equation-2d}, with $\beta=0$ and boundary conditions: $\kappa(i\ell)=\kappa((i+1)\ell)=\kappa_0$. Without loss of generality we consider a segment located in $0\le s \le \ell$. Although the exact analytic solution this nonlinear equation is available (see Sec. \ref{sec:plane}), an excellent approximation suitable for this purpose can be obtained by neglecting the cubic nonlinearity (Fig.~\ref{fig:spreading}C,D). With this simplification, Eq. \eqref{eq:euler-lagrange} admits a simple solution of the form:
\begin{equation}\label{eq:curvature}
\kappa(s)=\kappa_0 + \frac{\sigma}{4\alpha}\,s(s-\ell)\;.
\end{equation}
Eqs. \eqref{eq:curvature} and \eqref{eq:discontinuity} immediately allow us to derive a condition on the cell perimeter: $\ell=2k_{s}\Delta/\sigma$. Furthermore, the latter condition leads to a linear relation between traction force $T= 2 k_s \Delta$, and cell size :
\begin{equation}\label{eq:traction-length}
T=\sigma \ell\;,
\end{equation}
which is indeed observed in traction force measurements on large epithelial cells~\cite{Mertz:2012}.

To determine the end-point curvature $\kappa_{0}$, we use the turning tangents theorem for a simple closed curve \eqref{eq:turning-tangents}. This leads to following relation between local curvature and segment length, or equivalently traction force, at the adhesion sites :
\begin{equation}
\kappa_{0}=\frac{\sigma\ell^{2}}{24\alpha} + \frac{2\pi}{N_{A}\ell}
\end{equation}

Finally, to determine the optimal length of the cell segment $\ell$, we are going to make use of a remarkable geometrical property of the curve obtained from the solution of Eq.~\eqref{eq:euler-lagrange} with discrete adhesions: the fact of being a {\em curve of constant width} \cite{Gray:1997}. The width of a curve is the distance between the uppermost and lowermost points on the curve (see lower inset of Fig. \ref{fig:spreading}B). In general, such a distance depends on how the curve is oriented. There is however a special class of curves, where the width is the same regardless of their orientation. The simplest example of a curve of constant width is clearly a circle, in which case the width coincides with the diameter. A fundamental property of curves of constant width is given by the Barbier's theorem \cite{Gray:1997}, which asserts that the perimeter $L$ of any curve of constant width is equal to width $w$ multiplied by $\pi$: $L=\pi w$. As illustrated in Fig.~\ref{fig:spreading}D, this is confirmed by numerical simulations for low to intermediate values for contractility and stiffness. With our setting, the cell width is given by:
\begin{equation}\label{eq:width}
w = (R_{0}-\Delta)(1+\cos\pi/N_{A})+h(\ell/2)\;,
\end{equation}
where $h(s)=\int_{0}^{s}ds'\,\sin\theta(s')$ is the height of the curve above a straight line connecting subsequent adhesions and $\theta$ the turning angle. For small deflections $h$ can be approximated as:
\begin{equation}
h(s) \approx s(\ell-s)\left[\frac{\pi}{N_{A}\ell}-\frac{\sigma}{48\alpha}\,s(\ell-s)\right]\;.
\end{equation}
Using this together with Eq. \eqref{eq:width} and the Barbier's theorem with $L=N_{A}\ell$ allow us to obtain a quartic equation for the cell length: 
\begin{multline}\label{eq:traction}
\frac{N_{A}\ell}{\pi} = (1+\cos\pi/N_{A})\left(R_{0}-\frac{\sigma}{2k_{s}}\,\ell\right)\\[5pt]
+\frac{\ell}{4}\,\left(\frac{\pi}{N_{A}}-\frac{1}{192}\,\frac{\sigma\ell^{3}}{\alpha}\right)\;.
\end{multline}
Fig. \ref{fig:traction}A,B show plots of the traction \eqref{eq:traction-length} with $\ell$ determined by solving Eq. \eqref{eq:traction}. The plots support the experimental trend that traction increases monotonically with substrate stiffness $k_{s}$ before plateauing to a finite value for higher stiffnesses~\cite{Ghibaudo:2008,Mitrossilis:2009}. The plateau value increases with increasing contractility (Fig.~\ref{fig:traction}A). Traction force grows linearly with increasing contractility for $\sigma R_0^3/\alpha \ll 1$, before saturating to the value $2k_s R_0$ at large contractility $\sigma R_0^3/ \alpha \gg 1$, as shown in Fig.~\ref{fig:traction}B. 
\begin{figure}[t]
\centering
\includegraphics[width=\columnwidth]{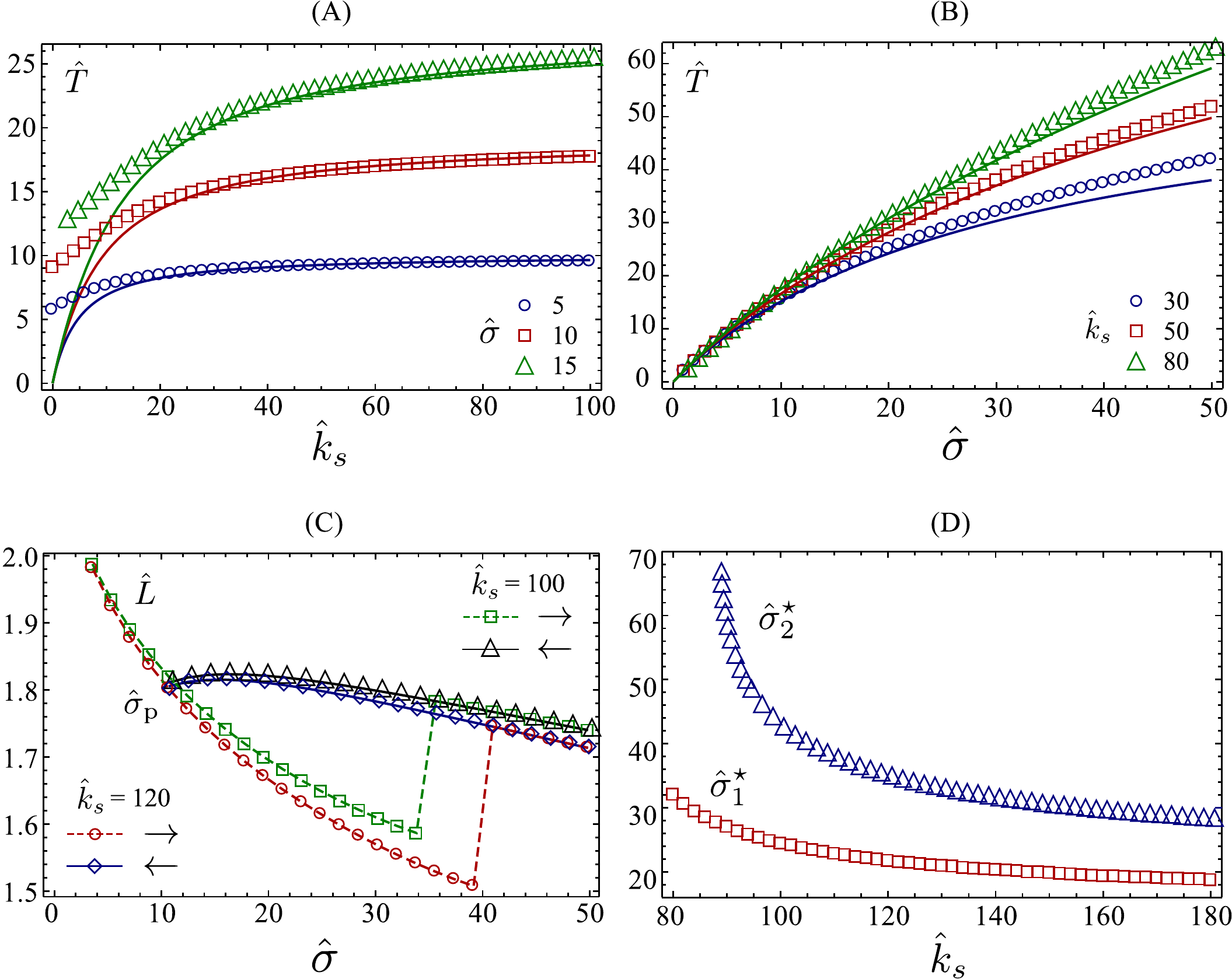}
\caption{\label{fig:traction}Traction force as a function of substrate stiffness (A) and contractility (B) obtained from a numerical minimization of a discrete analog of Eq. \eqref{eq:energy}. For convenience the various quantities have been rescaled as follows: $\hat{L}=L/R_{0}$, $\hat{T}=T R_{0}^{2}/\alpha$, $\hat{\sigma}=\sigma R_{0}^{3}/\alpha$ and $\hat{k}_{s}=kR_{0}^{3}/\alpha$. The solid curves denote the approximate traction values obtained from Eq. \eqref{eq:traction}. (C) Boundary length $\hat{L}$ obtained by increasing (squares) and then decreasing (triangles) the contractility for substrate stiffnesses $\hat{k}_{s}=100$ (green squares, black triangles) and $\hat{k}_{s}=120$ (red squares, blue triangles). The diagram shows bistability in the range $\hat{\sigma}_{\rm p} < \sigma < \sigma^{\star}_{2}$. (D) The critical contractility $\sigma^{\star}_{1}$ and $\sigma^{\star}_{2}$ as functions of substrate stiffness.}
\end{figure}
\begin{figure}[t]
\centering
\includegraphics[width=0.8\columnwidth]{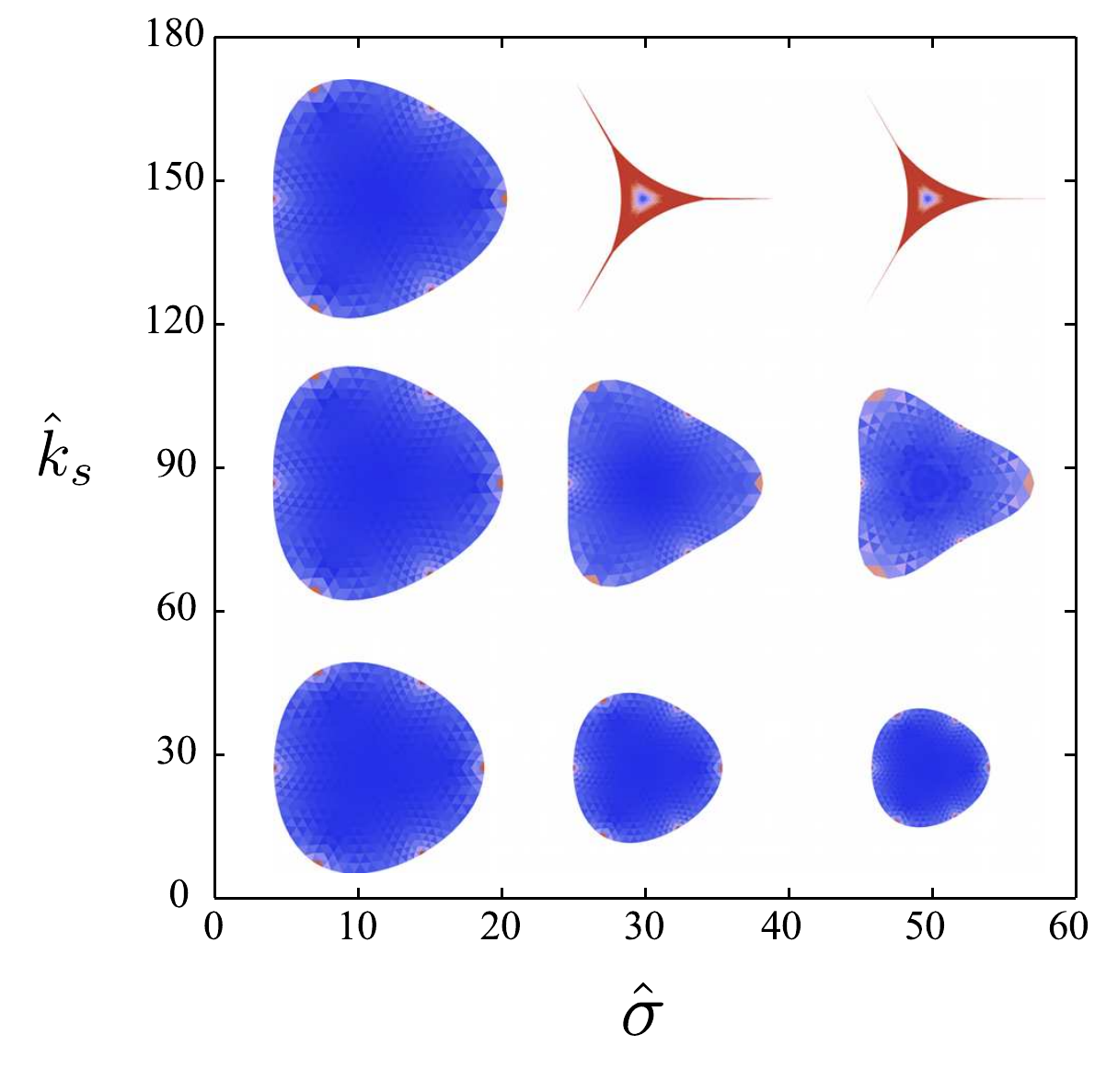}
\caption{\label{fig:phase}Phase diagram in plane of $\hat{\sigma}=\sigma R_{0}^{3}/\alpha$ and $\hat{k}_{s}=k_{s}R_{0}^{3}/\alpha$ showing optimal configuration obtained by numerical minimization of the energy \eqref{eq:energy} for $N_{A}=3$.}
\end{figure}%
\subsection{Inflections, cusps and protrusions}
For low to intermediate values of $\sigma$ and $k_s$, cell shape is convex and has constant width. Upon increasing $\sigma$ above a $k_{s}-$dependent threshold $\sigma^{\star}_{1}$, however, the cell boundary becomes inflected (see Fig. \ref{fig:phase} and upper inset of Fig. \ref{fig:spreading}B). Initially a region of negative curvature develops in proximity of the mid point between two adhesions, but as the surface tension is further increased, the size of this region grows until positive curvature is preserved only in a small neighborhood of the adhesion points. Due to the presence of local concavities, the cell boundary is no longer a curve of constant width. 

Upon increasing $\sigma$ above a further threshold value $\sigma^{\star}_{2}$, the inflected shape collapses giving rise to the star-shaped configurations shown in upper right corner of Fig.~\ref{fig:phase}. These purely concave configurations are made by arcs whose ends meet in a cusp. The cusp is then connected to the substrate by a protrusion consisting of a straight segment of that extends until the adhesion point rest position, so that $\Delta\approx 0$ (Fig. ~\ref{fig:schematic}C). The cell boundary becomes pinned at adhesion sites as a result of having to satisfy force-balance, Eq.~\eqref{eq:euler-lagrange}, and adhesion-induced boundary condition, Eq.~\eqref{eq:discontinuity}, while accommodating large contractile tensions at its neighbourhood. This results in spontaneous expansion in the cell perimeter. Unlike the previous transition from convex to non-convex shapes, this second transition occurs discontinuously and is accompanied by a region of bistability in the range $\sigma_{\rm p}<\sigma < \sigma^{\star}_{2}$. This is clearly visible in the hysteresis diagram in Fig.~\ref{fig:traction}C showing the optimal length obtained by numerically minimizing a discrete analog of Eq. \eqref{eq:energy} in a cycle and using as initial configuration the output of the previous minimization. The onset of bistability is regulated by substrate stiffness as shown in Fig.~\ref{fig:traction}D, with stiffer substrates promoting transition to cusps at lower $\sigma^{\star}_{2}$. Away from the protrusion, the curvature has still the form given in Eq. \eqref{eq:curvature}, with $\kappa_{0}=0$ so that the boundary is everywhere concave or flat and the bending moment $\bm{M}=2\alpha\kappa\bm{z}$ does not experience any unphysical discontinuity at the protrusions origin.

At $\sigma=\sigma_{{\rm p}}$ the protrusion have zero length and the corresponding configuration of the cell boundary form a special curve from which all the shapes having non-zero protrusion length can be constructed by means of the similarity transformation \eqref{eq:similarity} discussed in Sec. \ref{sec:buckled-shapes}. To see this let $\ell_{\rm p}$ be the protrusion length and let us set $\ell_{\rm p}=0$ at $\sigma=\sigma_{{\rm p}}$, so that the shape of the cell will be of the kind illustrated in Fig.~\ref{fig:schematic}B. Then given the surface tension $\sigma>\sigma_{\rm p}$ we calculate the scaling factor $\lambda=(\sigma_{\rm p}/\sigma)^{1/3}$, we rescale the reference curve by $\lambda$ and finally we fill the distance between the adhesion points and the cusps with straight segments of length $\ell_{\rm p}=R_{0}(1-\lambda)$ (since $R_{0}$ is the circumradius of the reference shape and $\lambda R_{0}$ that of the rescaled shape). This latter step, ultimately allows us to formulate a scaling law for the length of protrusions that can be tested in experiments:
\begin{equation}\label{eq:protrusion}
\ell_{\rm p}/R_{0} = 1-(\sigma_{\rm p}/\sigma)^{1/3}\;.	
\end{equation}
This transition from a smooth shape to a self-contacting shape with cusps is reminiscent of the post-buckling scenario of an elastic ring subject to a uniform pressure reviewed in Sec. \ref{sec:plane}, but unlike this case, where the system undergoes a continuous transition from a simple curve to a curve with lines of contact, here the transition is discontinuous along both the loading branch (increasing $k_{s}$) and the unloading branch (decreasing $k_{s}$). The transition has moreover a strong topological character since it involves a jump in the rotational index of the curve, whose total curvature after the transition becomes:
\[
\oint_{\partial M}ds\,\kappa = \pi(2-N_{A})\;,
\]
by virtue of Eq. \eqref{eq:turning-kinks}. Some further detail about the geometry of protrusions in this model can be found in Ref. \cite{Banerjee:2013}. It should be stressed, however, that our knowledge of this phenomenon is still very preliminary and the origin itself of this exotic instability (which in same extent is reminiscent of the {\em sulcification} instability in neo-Hookean materials \cite{Biot:1965,Hohlfeld:2011,Hohlfeld:2012}) still unclear. One of the fundamental aspect that distinguishes the Contractile Film Model form classical elasticity relies on the fact that also the perimeter is not hardly constrained, but only subject to a soft constraint by mean of the adhesion springs. The length of an elastic object affects its overall flexibility (i.e. long filaments are floppy and easy to bend, while short filaments are stiff), thus, when the effective surface tension is increased, the whole cell boundary becomes shorter and stiffer. Because stiff materials are difficult to bend, but easy to break, a possible interpretation could be the following. For sufficiently large adhesion, increasing the surface tension has the effect of bending and stiffening the cell boundary in proximity of the adhesion sites, until, above a certain surface tension, the cell boundary is too stiff to continue bending and fractures. The cracks are localized at the adhesion points, where the curvature initially focuses, giving rise to the cusps observed in the simulations. However, a thorough understanding of this phenomenon remains a challenge for the future.

\section{\label{sec:conclusions}Conclusions}

In this article we reviewed some of the fundamental mathematics and physics of softly constrained films. Starting from the original 19th century query of finding the shape of a two-dimensional elastic ring subject to a uniform pressure \cite{Levy:1884}, we retraced the developments of this concept until the most recent analytical and experimental progress \cite{Vassilev:2008,Djondjorov:2011,Mora:2012,Giomi:2012} and we examined an application to the important problem of cell adhesion. While inspired by the observation of the mundane, the subject has received renewed interest by a heterogeneous community of physicists, applied mathematicians and engineers and is now a powerful prototype problem with potential applications to various physical systems whose geometry is softly constrained. Applications to biological tissues and cells are by far the most promising. Although it relies only on a few fundamental physical mechanisms, compared to the extraordinarily complex orchestration of the cell machinery, the Contractile Film Model has provided some interesting insight on the geometry of adherent cells, including the experimentally observed transition from convex to concave geometry through the formation of cusps and protrusions. While a thorough understanding of the latter phenomena requires a deeper investigation, the scenario suggested by these preliminary results is very promising: is a whole new class of biologically relevant instabilities and morphological transitions lurking beyond geometrical constraints?

\acknowledgments
I am indebted to L. Mahadevan, who inspired and contributed to the work reviewed in Sec. \ref{sec:ep}, and to Shiladitya Banerjee, who contributed to the work reviewed in \ref{sec:cell}. I am also grateful to Serge Mora for sharing with me the beautiful experimental image shown in Fig. \ref{fig:mora}. This work is supported by SISSA mathLab.

\bibliographystyle{unsrtnat}
\bibliography{softly-constrained-films} 
\end{document}